\documentclass[twocolumn,aps,prb,showpacs,preprintnumbers,amsmath,amssymb]{revtex4-1}
\usepackage{natbib}
\usepackage{graphicx}
\usepackage{bm}
\usepackage{gensymb}
\usepackage{color}
\begin{document}
\title{Polaron dynamics and decoherence
 in an interacting two-spin system coupled to optical phonon environment}
\author{Amit Dey$^{1}$}
\author{Sudhakar Yarlagadda$^{1,2}$}
\affiliation{${^1}$TCMP Div., Saha Institute of Nuclear Physics,
Kolkata, India}
\affiliation{${^2}$Cavendish Lab, Univ. of Cambridge, Cambridge, UK}
\date{\today}

\begin{abstract}
{{ We study two anisotropically interacting spins 
 coupled to optical phonons; we restrict our analysis to the regime of strong coupling
to the environment, to the antiadiabatic region, and to the subspace with zero value for
$S^z_T$ (the z-component of the total spin). In the case where each spin is coupled to a 
different phonon bath, we assume that the system and the environment are initially
uncorrelated (and form a simply separable state) in the polaronic frame of reference. By}} analyzing the {polaron} dynamics through
non-Markovian quantum master equation, we find that the system manifests
a small amount of decoherence that decreases both with increasing non-adiabaticity and with enhancing strength of coupling; 
whereas, under the Markovian approximation, the polaronic system exhibits a decoherence free behavior. 
For the situation where both spins are coupled to the same phonon bath, we also show that
the system is decoherence free in the subspace where $S^z_T$ 
is fixed.  
To suppress decoherence through quantum control, we employ a train of pi pulses and demonstrate that 
unitary evolution of the system can be retained.{{We propose realization of a weakly decohering charge qubit
from an electron in an oxide-based (tunnel-coupled) double quantum dot system}}. 
\end{abstract}
\maketitle
\section{Introduction}
 Quantum computation relies on the very quantum features of systems, namely, entanglement and quantum superposition. Due to the ubiquitous
presence of system-environment interactions, the quantum coherence is degraded;  on total destruction of the quantum superposition,
 classicality emerges. 

The protection of quantum states against decoherence due to interaction with the environment
is  essential to the development of quantum computational architecture. 
In this regard quantum robustness strategies have been developed
that are passive [such as decoherence free subspaces (DFS)]
as well as active [such as quantum Zeno effect (QZE) and its variants].
A DFS is a subspace that is both invariant under the action of the system
Hamiltonian and also is spanned by degenerate eigenvectors of the
system operators coupling to the environment \cite{zanardi,whaley1,whaley2}.
QZE is the arresting of decoherence in
 a quantum system through continuous measurements
\cite{Sudarshan,kofman,facchi}.  Continuous measurement can be approximately achieved  
if one makes a series of instantaneous measurements with the time interval between consecutive ones being sufficiently small.
Mathematically reminiscent of the QZE is the  
dynamical decoupling through bang-bang control \cite{Lidar,Viola, Morton1}.
Now, 
in this connection, one 
can think of suppressing the nonunitary evolution of an open system by using an external fast driving field.  
Earlier, strategies have been proposed to suppress decoherence by perturbing the system (with perturbation being experimentally controllable) 
 in such a manner that the environment can not follow the change of system states anymore; this means, if the perturbation acts on the 
system much faster than the environmental response time, the system can be effectively decoupled from the environmental fluctuations and thus decoherence 
is diminished \cite{Viola1,Ulrich1}.

Using control techniques, achievement 
of long coherence times has been reported in GaAs and silicon based double quantum dots where the qubit information is encoded in singlet-triplet states
involving two spins 
\cite{nat1, petta, yacoby, Maune, Barthel}.
 Moreover, the two-spin states with 
fixed z component of the total spin ($S^z_T=0$) are immune to collective decoherence 
 and using Heisenberg exchange interaction 
universal quantum computation has been proposed for this subspace \cite{Levy}.

In this paper, we consider an anisotropic Heisenberg interaction between
two spins with zero value for 
$S^z_T$.
 Each spin is taken to interact strongly with its individual optical phonon bath. 
{For studying the dynamics and decoherence,
 the polaronic (Lang-Firsov \cite{lang} transformed) frame of reference
is chosen with polaronic (dressed) basis used for input and output \cite{lidar}; unlike the
interaction in the original frame of reference, the interaction term is weak in the transformed polaronic frame
enabling one to carry out perturbation theory.}  Since both preparation and measurement
can be done in the dressed basis, 
we analyze decoherence  in the transformed
polaronic frame of reference; thus {\em we seek the possibility of maintaining quantum information encoded in dressed 
 spin (polaron) basis.}
We study the dynamics of the reduced spin system using quantum master equation approach
when initially
the system and the environment form a simply separable state. 
 We find that the off-diagonal 
 matrix elements indicate small decoherence under non-Markovian dynamics; on the other hand,
Markovian quantum master equation yields decoherence free behavior. We also show that  decoherence (in the non-Markovian case)
can be suppressed using bang-bang control and unitary evolution of the spin states 
 can be achieved.

For the case 
  where both the spins interact with the same phonon environment (through a global coupling), the system is decoherence free only in the subspace
where $S^z_T$
 is a fixed constant. 

{{One can produce new oxide-based devices by exploiting their tunability, rich physics, and coupling between
  the various degrees of freedom (such as charge, lattice, spin, etc.).
Here, we also propose that oxide-based double quantum dots with only one electron (tunneling
between the dots) can be regarded as a qubit with little decoherence. {Strong coupling between electrons and phonons
occurs quite commonly in transition metal oxides such as manganites \cite{lanzara,bishop,pbl}}.
Oxides, due to the smaller extent of the wavefunction, present a promising alternative to silicon for meeting the  demands of miniaturization.}}

 \section{Model for local phonon coupling}
 Here we consider the model given by
 \begin{eqnarray}
 \label{model}
  \nonumber H &=& H_s + H_{B} + H_I ,
\end{eqnarray}
where $H_s$ is the system Hamiltonian describing the anisotropic Heisenberg interaction between two spins
\begin{eqnarray} 
\label{heisenberg}  
H_s = J_\lVert S_{1}^{z}S_{2}^{z} {{-}} J_\perp (S_{1}^{x}S_{2}^{x}+S_{1}^{y}S_{2}^{y}) ,
\end{eqnarray}
{$H_{B}$ corresponds to the phononic bath
\begin{eqnarray}
H_{B} = \omega \sum_{i=1,2} a_{i}^{\dagger} a_i ,
\end{eqnarray}
}and {$H_I$ represents the local (i.e., on site) spin-bath interaction}
\begin{eqnarray}
H_I = 
   g\omega\sum_{i=1,2} S_{i}^{z}( a_i + a_{i}^{\dagger}).
 \end{eqnarray}
In the above equation, $\omega$ is the optical phonon frequency, $g$ is the spin-phonon coupling parameter (with $g^2 \gg 1$), and $a_i$ is the phonon 
destruction operator at site $i$.  {As mentioned in Ref. \onlinecite{sdadsy}, for manganites $6 < g^ 2 < 10$. 
{{ The actual bath and interaction terms involve various momentum modes and have the form
$H_{B} =  \sum_{i,k} \omega_k a_{i,k}^{\dagger} a_{i,k}$ and
$ \sum_{i,k} g_k \omega_k S_{i}^{z}( a_{i,k} + a_{i,k}^{\dagger})$ respectively. Initially,
for simplicity, we ignore the wavenumber dependence of the coupling and the frequency
and take $g_k=g$ and $\omega_k=\omega$; the more general case, involving wavenumber dependence, will be analyzed later in Appendix C.}}
Furthermore, the unperturbed Hamiltonian $H_0$ is given by
\begin{equation}
  H_0 = H_s + H_{B}.
 \end{equation}
Here we are dealing with $S=\frac{1}{2}$ system; consequently,
the four eigenstates of $H_s$ are the  triplet states
 \{ $|\uparrow \uparrow\rangle ,|\downarrow \downarrow\rangle, 
\frac{1}{\surd 2}(|\uparrow \downarrow\rangle + |\downarrow \uparrow\rangle)$\} 
and the singlet state \{$\frac{1}{\surd 2}(|\uparrow \downarrow\rangle - |\downarrow \uparrow\rangle)$\}.
 
\section{Lang-Firsov transformation}
In the non-adiabatic ($\frac{J_\perp}{\omega}\leq 1$) and the strong coupling ($g^2 \gg 1$) regimes, we make use of the well-known 
Lang-Firsov ($L$) transformation given by \cite{lang}
\begin{equation}
 H^{L}=e^{S}He^{-S},
\end{equation}
where $S= -\sum_{i}g S^z_i (a_i - a_{i}^{\dagger})$. Therefore the total Hamiltonian in the Lang-Firsov frame is written as
\begin{eqnarray}
 \nonumber H^{L}&=& J_\lVert S_{1}^{z}S_{2}^{z} {{-}} \frac{J_\perp e^{-g^2}}{2} (S_{1}^{+}S_{2}^{-}+S_{2}^{+}S_{1}^{-})\\
  &&+ \omega \sum_{i=1,2} a_{i}^{\dagger} a_i {{-}} \frac{1}{2} [J_\perp ^+ S_{1}^{+}S_{2}^{-} + J_\perp ^- S_{2}^{+}S_{1}^{-} ],
\label{HLF}
\end{eqnarray}
 where 
 \begin{eqnarray}
 J_\perp ^{\pm} &=& J_\perp e^{\pm g[(a_2 - a_{2}^{\dagger})-(a_1 - a_{1}^{\dagger})]} -  J_\perp e^{-g^2},
\end{eqnarray}
and
\begin{eqnarray}
 J_\perp e^{-g^2} &=&\langle J_\perp e^{\pm g[(a_2 - a_{2}^{\dagger})-(a_1 - a_{1}^{\dagger})]} \rangle_{T=0},
 \end{eqnarray}
 {where the average is taken over the phonon ground state.}
In the Lang-Firsov frame, we recognize that
\begin{equation}
 H_s^{L}= J_\lVert S_{1}^{z}S_{2}^{z}  {{-}}\frac{J_\perp e^{-g^2}}{2} (S_{1}^{+}S_{2}^{-}+S_{2}^{+}S_{1}^{-}),
\label{hsl}
\end{equation}
as the system Hamiltonian and this describes spins coupled to the same mean phonon field; it is interesting to note that both $H_s$ and $H_s^{L}$ have the
 {{singlet as the excited state and the $S^z_T=0$ triplet state as the ground state}}.
Furthermore, as will be shown later, the effective Hamiltonian (obtained from Schrieffer-Wolff
transformation, i.e.,  second-order perturbation theory involving averaging
over the ground state of phonons \cite{sw,sahinur}) also has the same ground state and excited state as $H_s^{L}$ [see Eq. (\ref{dom_2})].

{Now}, for $H_s^{L}$, in the strong coupling ($g^2 \gg 1$) and the non-adiabatic ($\frac{J_\perp}{\omega}\leq 1$) regimes
 the energy difference between the singlet and the $S^z_T =0 $ triplet state 
is $J_\perp e^{-g^2} \ll \omega$. The interaction Hamiltonian $H_I^{L}$, describing the spins coupled to the local phonon 
fluctuations around the mean field, is given by
\begin{eqnarray}
 H_I^{L}&=& {{-}}\frac{1}{2} [J_\perp ^+ S_{1}^{+}S_{2}^{-} + J_\perp ^- S_{2}^{+}S_{1}^{-} ] ,
\label{hil}
\end{eqnarray}
whereas the environment Hamiltonian $H_{B}^{L}$, consisting of 
displaced harmonic oscillators, is expressed as
\begin{eqnarray}
 H_{B}^{L} &=& \omega \sum_{i=1,2} a_{i}^{\dagger} a_i.
\label{hbl}
\end{eqnarray}
Furthermore, the unperturbed Hamiltonian $H_0^{L}$ is identified
as
\begin{eqnarray}
 H_0^{L}=H_s^{L}+H_B^{L} .
\end{eqnarray}
Here, it should be mentioned that the mean field term $H_s^{L}$ involves controlled degrees of freedom and hence produces unitary
 evolution for the system; whereas the interaction Hamiltonian $H_I^{L}$ contains {numerous} or uncontrolled environmental degrees of freedom
and consequently has the potential for producing decoherence. Now, at this stage, we like to analyze decoherence due 
to these local environmental fluctuations from a dynamical perspective.
\section{Dynamical analysis for local coupling}
The two-spin system is 
exposed to the environment (with which it interacts) thereby making it an open system; we would like to know  the reduced system dynamics
for this open system.
Here, we deal with non-Markovian master equation and try to obtain the off-diagonal elements 
of the reduced 
density matrix for the system defined by 
{$\rho_s(t) \equiv Tr_B\left[\rho_T(t)  \right]$ where the bath degrees} of freedom 
are traced out from the total 
system-environment density matrix $\rho_T(t)$; the decay of the off-diagonal elements of $\rho_s(t)$ signifies quantum decoherence.
 We assume that the {
{initial total density matrix is simply separable, i.e., $\rho_T(0) = \rho_s(0) \otimes B_0$, where 
$B_0=\sum_{n_1,n_2}|n_1,n_2 \rangle_{ph}~\!_{ph}\langle n_1,n_2|e^{-\beta \omega_{n_1,n_2}}/Z$ 
is the initial thermal density matrix of the bath under thermodynamic equilibrium.  The state
$|n_1,n_2 \rangle_{ph}$ represents a phonon state with $n_1$ and $n_2$ being the
phonon occupation numbers at sites $1$ and 
$2$; furthermore,
$H_{B}^{L}|n_1,n_2\rangle_{ph} = \omega_{n_1,n_2} |n_1,n_2 \rangle_{ph}=(n_{1}+n_2) \omega |n_1,n_2 \rangle_{ph}$ and $Z$ is the partition function 
for the environment.} {From here on we use  
$|n\rangle_{ph} \equiv |n_1,n_2 \rangle_{ph}$, $\sum_{n} \equiv \sum_{n_1,n_2}$, and $\omega_n \equiv \omega_{n_1,n_2}  = \omega (n_1+n_2)$.}}
Here it should be mentioned that 
 states can be prepared and measured in the dressed (polaronic) basis and that this dressed
basis can be used for input and output \cite{lidar}.
We make our analysis in the Lang-Firsov frame because in this frame the spins 
can be initially decoupled from the phonon environment 
({although in the original frame of reference the spins 
are entangled with their local environment}). Moreover, 
since the interaction in the Lang-Firsov frame 
is weak (compared to that in the original frame of reference),  perturbation theory can be applied. Now, starting with the von Neumann 
equation in the interaction picture 
\begin{eqnarray}
 \frac{d \tilde{\rho}_T (t)}{dt} = -i [\tilde{H}_I^L(t), \tilde{\rho}_T(t)] ,
\end{eqnarray}
and using the time-convolutionless projection operator technique, one obtains the 
second order time-convolutionless (TCL) non-Markovian quantum master equation \cite{open} given by
\begin{eqnarray}
\!\!\!\! {\frac{d \tilde{\rho}_s(t)}{dt} 
= - \int_0^t d\tau Tr_B[\tilde{H}_I^L(t),[\tilde{H}_I^L(\tau), \tilde{\rho}_s(t) \otimes B_0]] .}
\label{mas2}
\end{eqnarray}
Here, expressed in the interaction picture representation,  $\tilde{H}_I^L(t) = e^{iH_0^{L} t} H_I^L e^{-iH_0^{L}t} $ is the interaction Hamiltonian
and $\tilde{\rho}_T (t) = e^{iH_0^{L}t} \rho_T(t) e^{-iH_0^{L}t} $ 
is the total 
density matrix in the Lang-Firsov frame.
On expanding the double commutator and using the complete set of phonon states 
$\sum_{m}|m \rangle_{ph}~\! _{ph}\langle m|=I$, one can write  equation (\ref{mas2}) as
\begin{widetext}
\begin{eqnarray}
\frac{d \tilde{\rho}_s(t)}{dt} &=& - \frac{1}{{Z}}\sum_{n,m} \int_0^t d\tau \left[~ _{ph}\langle n |
 \tilde{H}_I^L(t)| m \rangle_{ph}~ _{ph}\langle m | \tilde{H}_I^L(\tau)| n \rangle_{ph} \tilde{\rho}_s(t) e^{-\beta \omega_{n}} \right. 
\nonumber \\ 
 &&~~~~~~~~~~~~~~~~~~~ - {~_{ph}}\langle n| \tilde{H}_I^L(t)|m\rangle_{ph}\tilde{\rho}_s(t) {_{ph}}\langle m | \tilde{H}_I^L(\tau) |n\rangle_{ph}e^{-\beta \omega_{m}} \nonumber \\
 && ~~~~~~~~~~~~~~~~~~~- {~_{ph}}\langle n| \tilde{H}_I^L(\tau)| m\rangle_{ph} \tilde{\rho}_s(t) {_{ph}}\langle m| \tilde{H}_I^L(t) |n\rangle_{ph}e^{-\beta \omega_{m}} 
 \nonumber \\
&& ~~~~~~~~~~~~~~~~~~~
 + \left .
~ \tilde{\rho}_s(t)  _{ph}\langle n |
 \tilde{H}_I^L(\tau)| m \rangle_{ph}~ _{ph}\langle m | \tilde{H}_I^L(t)| n \rangle_{ph} e^{-\beta \omega_{n}} 
 \right] .\nonumber \\
\label{mas}
\end{eqnarray}

We are interested in the dynamics at ${\rm T=0~K}$. The first term in equation (\ref{mas}) (at ${\rm T=0~K}$) can be expressed as 
\begin{eqnarray}
 \nonumber &&_{ph}\langle 0 |\tilde{H}_I^L(t)| m\rangle_{ph}~ _{ph}\langle m | \tilde{H}_I^L(\tau)| 0 \rangle_{ph} \tilde{\rho}_s(t)
\\ &&~~~~~~~~~=
  \sum_{\varepsilon, \varepsilon^{\prime}, \varepsilon^{\prime \prime}} \left [ |\varepsilon \rangle {\!\langle} \varepsilon| ~ 
  _{ph}\langle 0 | H_I^L |m\rangle_{ph} 
|\varepsilon^{\prime} \rangle {\!\langle} \varepsilon^{\prime} |~ _{ph}\langle m | H_I^L |0\rangle_{ph}
 |\varepsilon^{\prime \prime} \rangle {\!\langle} \varepsilon^{\prime\prime} | 
e^{i[(\varepsilon-\varepsilon^{\prime})t+(\varepsilon^{\prime}-\varepsilon^{\prime\prime})\tau]}   \right]
 \tilde{\rho}_s(t)
 e^{-i \omega_{m}(t-\tau)} ,
\label{time1}
\end{eqnarray}
where $\sum_{\varepsilon} |\varepsilon \rangle {\!\langle} \varepsilon|=I$ is the completeness relation in terms of the eigenstates
defined as
$H_s^{L}|\varepsilon \rangle=\varepsilon|\varepsilon \rangle$. Now,
we first note that, for a fixed $S_T^z$, only the hopping term contributes to the excitation gap. Then,
 for a fixed $S_T^z$ ($=0$), the condition $J_\perp e^{-g^2} \ll \omega$ 
implies $\omega_{m} \gg |\varepsilon - \varepsilon^{\prime}|$ and
$ \omega_{m} \gg |\varepsilon^{\prime} -\varepsilon^{\prime \prime}|$;
hence, in equation (\ref{time1}), we can take 
\begin{eqnarray}
{e^{i[(\varepsilon-\varepsilon^{\prime})-\omega_{m}]t} \approx e^{-i\omega_{m} t}} ,
\label{eps_om1}
\end{eqnarray}
 and
\begin{eqnarray}
{e^{i[(\varepsilon^{\prime}-\varepsilon^{\prime \prime})+\omega_{m}]\tau} \approx e^{i\omega_{m} \tau}} .
\label{eps_om2}
\end{eqnarray}
In Appendix B, we consider the effect of the phase factor due to not ignoring $ |\varepsilon_s -\varepsilon_t|$ compared to $\omega$.
Using the approximations given in Eqs. (\ref{eps_om1}) and (\ref{eps_om2}), we can write
\begin{eqnarray}
\label{te}
 _{ph}\langle 0 |\tilde{H}_I^L(t)| m \rangle_{ph}~ _{ph}\langle m | \tilde{H}_I^L(\tau)| 0 \rangle_{ph} ~\tilde{\rho}_s(t) 
 = {_{ph}\langle} 0 |{H}_I^L| m \rangle_{ph}~ _{ph}\langle m | {H}_I^L| 0 \rangle_{ph} ~\tilde{\rho}_s(t) 
 e^{-i\omega_{m}(t-\tau)}.
\end{eqnarray}
At ${\rm T=0~K}$, using the same reasoning for all the four terms in equation (\ref{mas}), we obtain 
\begin{eqnarray}
\label{zero}
  \frac{d \tilde{\rho}_s(t)}{dt}&=& - \int_0^t d\tau \Big[\sum_{m} [ |_{ph}\langle 0 | H_I^L |m\rangle_{ph}|^2
 ~\tilde{\rho}_s(t)  e^{-i\omega_{m}(t-\tau)} \nonumber \\ 
 &&~~~~~~~~~~~~~~~~~\nonumber + \tilde{\rho}_s(t) ~ |_{ph}\langle 0 | H_I^L |m\rangle_{ph}|^2 
  e^{i\omega_{m} (t-\tau)}] \\
&&~~~~~~~~~~~~ - \sum_{n}  [ _{ph}\langle n | H_I^L |0\rangle_{ph}  \tilde{\rho}_s(t) _{ph}\langle 0|
 H_I^L|n\rangle_{ph} e^{i\omega_{n}(t-\tau)} \nonumber\\ 
&&~~~~~~~~~~~~~~~~~~~~+  {~_{ph}}\langle n | H_I^L |0\rangle_{ph}  \tilde{\rho}_s(t) _{ph}\langle 0| H_I^L|n\rangle_{ph}
 e^{-i\omega_{n} (t-\tau)}] \Big].
 \end{eqnarray}
 \end{widetext}
 On calculating the matrix element $_{ph}\langle 0 | H_I^L |m\rangle_{ph}$ we have
\begin{eqnarray}
\nonumber && _{ph}\langle 0 | H_I^L | m\rangle_{ph} = {_{ph}}\langle 0,0 | H_I^L | m_1,m_2\rangle_{ph} 
\nonumber \\
&& = {{-}}  \frac{J_\perp e^{-g^2}}{2} \frac{g^{m_1 + m_2}}{\sqrt{m_1 !  m_2 !}} 
\left ( (-1)^{m_1} S_{1}^{+}S_{2}^{-}+ (-1)^{m_2} S_{2}^{+}S_{1}^{-} \right ), 
\nonumber \\
\label{correction}
\end{eqnarray}
where $m_1$ and $m_2$ are not zero simultaneously. Moreover,
\begin{equation}
  _{ph}\langle 0 | H_I^L | 0 \rangle_{ph}=0.
  \label{1order}
\end{equation}
We are interested only in the states with z-component of the total spin $S_T^z =0$, i.e., the singlet state \\
 $|\varepsilon_s\rangle = 
\frac{1}{\surd 2}(|\uparrow \downarrow\rangle - |\downarrow \uparrow\rangle)$ 
 and the triplet state
 $|\varepsilon_t\rangle =\frac{1}{\surd 2}(|\uparrow \downarrow\rangle + |\downarrow \uparrow\rangle)$.
 By taking the matrix element with respect to $|\varepsilon_s\rangle$ and $|\varepsilon_t\rangle$ on both sides of equation (\ref{zero}), we get
 \begin{eqnarray}
  &&\!\!\!\!\!\!\! \frac{d \langle \varepsilon_s | \tilde{\rho}_s(t)|\varepsilon_t\rangle }{dt} \nonumber \\
  &&\!\!= -K \Bigg[ \sum^\prime_{m_1,m_2}~C_{m} 
  \langle \varepsilon_s | \tilde{\rho}_s(t)|\varepsilon_t\rangle \frac{\sin(\omega_{m}t)}{\omega_{m}}  \nonumber \\
 &&~~~~~~~  + \sum_{|m_1- m_2| = {\rm odd}} C_{m} 
  \langle \varepsilon_t | \tilde{\rho}_s(t)|\varepsilon_s\rangle \frac{\sin(\omega_{m}t)}{\omega_{m}} \nonumber \\
  &&~~~~~~~ + \sum^{\prime}_{|m_1- m_2| = {\rm even},0} C_{m} 
  \langle \varepsilon_s | \tilde{\rho}_s(t)|\varepsilon_t\rangle \frac{\sin(\omega_{m}t)}{\omega_{m}}\Bigg],
  \label{sz0}
 \end{eqnarray}
where the $\prime$ in $\sum^\prime$ implies the case $m_1=m_2=0$ is excluded in the summation;
$\sum_{|m_1- m_2| = {\rm even},0}$ ($\sum_{|m_1-m_2] = {\rm odd}}$)
represents 
sum over all even and 0 (odd) values of $|m_1- m_2|$;
 $\omega_{m}  = \omega (m_1+ m_2) $;  
$K\equiv \frac{J_\perp^2}{2}e^{-2g^2}$; and $C_m\equiv C_{m_1, m_2} \equiv\frac{g^{2(m_1+m_2)}}{m_1! m_2 !}$.
The above equation (\ref{sz0}) and its complex conjugate yield the solutions:
\begin{eqnarray}
  &&\langle \varepsilon_s | \rho_s(t)|\varepsilon_t\rangle e^{i(\varepsilon_s-\varepsilon_t)t} \nonumber \\
  &&= 
  \frac{1}{2}\Bigg[ \langle \varepsilon_s | \rho_s(0)|\varepsilon_t\rangle \Big\{\exp[-2K D_{\rm all}(t)] \nonumber \\ 
  &&~~~~~~~~~~~~~~~~~~~~~~~~+\exp[-2K D_{\rm even}(t)] \Big\} \nonumber \\ 
  &&~~~~~~~+\langle \varepsilon_t | \rho_s(0)|\varepsilon_s\rangle \Big\{\exp[-2K D_{\rm all}(t)] \nonumber \\ 
  &&~~~~~~~~~~~~~~~~~~~~~~~~~~~ -\exp[-2K D_{\rm even}(t) ] \Big\} 
\nonumber  \Bigg], \\
\label{sol_offdiag1}
\end{eqnarray}
and
\begin{eqnarray}
  &&\langle \varepsilon_t | \rho_s(t)|\varepsilon_s\rangle e^{i(\varepsilon_t-\varepsilon_s)t} \nonumber \\
  &&=
  \frac{1}{2}\Bigg[ \langle \varepsilon_t | \rho_s(0)|\varepsilon_s\rangle \Big\{\exp[-2K D_{\rm all}(t)] \nonumber \\ 
  &&~~~~~~~~~~~~~~~~~~~~~~~~ +\exp[-2K D_{\rm even}(t)] \Big\} \nonumber \\ 
  &&~~~~~~~+\langle \varepsilon_s | \rho_s(0)|\varepsilon_t\rangle \Big\{\exp[-2K  D_{\rm all}(t)] \nonumber \\ 
  &&~~~~~~~~~~~~~~~~~~~~~~~~~~~ -\exp[-2KD_{\rm even}(t)] \Big\} \nonumber \Bigg] , \\
  \label{sol_offdiag2}
\end{eqnarray}
{where we use the notation}
{
\begin{eqnarray}
  D_{\rm all}(t)&=&\sum^{\prime}_{m_1, m_2}C_{m} 
  \frac{(1-\cos(\omega_{m}t))}{\omega_{m}^2}, \\
  D_{\rm even}(t)&=&\sum^{\prime}_{|m_1- m_2| = {\rm even},0} C_{m} 
  \frac{(1-\cos(\omega_{m}t))}{\omega_{m}^2}.
\nonumber \\
\end{eqnarray}}
Thus, we see that the above solutions show some amount of decoherence for non-Markovian dynamics.
By employing the same procedure as above, on defining $D_{\rm odd} \equiv D_{\rm all} -D_{\rm even}$,
we obtain the following diagonal elements:
\begin{eqnarray}
 &&\langle \varepsilon_s | \rho_s(t)|\varepsilon_s\rangle \nonumber \\
 &&= 
  \frac{1}{2}\Bigg[ \langle \varepsilon_s | \rho_s(0)|\varepsilon_s\rangle \Big\{1
  +\exp[-2K D_{\rm odd}(t) ] \Big\} \nonumber \\ 
  &&~~~~~~~+\langle \varepsilon_t | \rho_s(0)|\varepsilon_t\rangle \Big\{1
  -\exp[-2KD_{\rm odd}(t)] \Big\} \nonumber \Bigg],
 \\
\label{sol_diag1}
\end{eqnarray}
and
\begin{eqnarray} 
 && \langle \varepsilon_t | \rho_s(t)|\varepsilon_t\rangle \nonumber \\
 &&= \frac{1}{2}\Bigg[ \langle \varepsilon_t | \rho_s(0)|\varepsilon_t\rangle \Big\{1
  +\exp[-2KD_{\rm odd}(t)] \Big\} \nonumber \\ 
  &&~~~~~~~+\langle \varepsilon_s | \rho_s(0)|\varepsilon_s\rangle \Big\{1
  -\exp[-2KD_{\rm odd}(t)] \Big\} \nonumber \Bigg]. \\
  \label{sol_diag2}
\end{eqnarray}
{{Before proceeding further, we would like to mention
that the above analysis is valid in the regime $k_B T/\omega <<1$; the finite temperature
case (i.e., $k_B T/\omega \gtrsim 1$) needs additional considerations and will be dealt with elsewhere}.

{{ For deriving the long-time behavior of the density matrix elements, we use the mathematical trick}} 
\begin{eqnarray}
 &&\int_0 ^\infty dt \frac{\sin(\omega_{m}t)}{\omega_{m}} \nonumber \\
 &&= \int_0 ^\infty dt \frac{e^{i\omega_{m}t}-e^{-i\omega_{m}t}}
 {2i\omega_{m} } \nonumber \\
 &&= \frac{1}{2i\omega_{m}} \Big[\int_0 ^\infty dt~ e^{i(\omega_{m} + i\eta) t} - \int_0 ^\infty dt~ e^{-i(\omega_{m} - i\eta) t} \Big]
 \nonumber \\
 &&= \frac{1}{\omega_{m}^2}  ,
 \label{trick}
 \end{eqnarray}
where $\eta \rightarrow +0$.
On considering the initial density matrix elements to be real, for large values of $g^2$, their long-time form is given by
(with details of derivation in Appendix A) 
\begin{eqnarray}
\label{longtime}
  |\langle \varepsilon_s | \rho_s(t)|\varepsilon_t\rangle| \Bigg|_{t\rightarrow \infty}  &=&
{\langle \varepsilon_s | \rho_s(0)|\varepsilon_t\rangle}~
  \exp\Big[ -\frac{\gamma^2}{4g^2} \Big] , \\
  \label{longtime1}
  \langle \varepsilon_s | \rho_s(t)|\varepsilon_s\rangle \Bigg|_{t\rightarrow \infty} &=& \frac{1}{2}  \langle \varepsilon_s | \rho_s(0)|\varepsilon_s\rangle
  \Big\{1+\exp\Big[-\frac{\gamma^2}{8g^2} \Big]\Big\} \nonumber \\
  &&+ \frac{1}{2}  \langle \varepsilon_t | \rho_s(0)|\varepsilon_t\rangle
  \Big\{1-\exp\Big[-\frac{\gamma^2}{8g^2} \Big]\Big\} ,\nonumber \\
\end{eqnarray}
and
\begin{eqnarray}
  \langle \varepsilon_t | \rho_s(t)|\varepsilon_t\rangle \Bigg|_{t\rightarrow \infty} &=& \frac{1}{2}  \langle \varepsilon_s | \rho_s(0)|\varepsilon_s\rangle
  \Big\{1-\exp\Big[-\frac{\gamma^2}{8g^2} \Big]\Big\} \nonumber \\
  &&+ \frac{1}{2}  \langle \varepsilon_t | \rho_s(0)|\varepsilon_t\rangle
  \Big\{1+\exp\Big[-\frac{\gamma^2}{8g^2} \Big]\Big\},\nonumber \\
\label{longtime3}
  \end{eqnarray}
  {where the parameter $\gamma=\frac{J_\perp}{g \omega}$.}
{{Here, it is important to recognize that the density matrix
does not take the form given by the Boltzmann distribution at large time because
the spins do not exchange any energy with the environment (as follows from $\varepsilon_s-\varepsilon_t= J_{\perp} e^{-g^2} \ll \omega$);
consequently, there are no exponential decay terms.
 }}
\begin{figure}
\centerline{\includegraphics[angle=90,angle=90,angle=90,width=3.2in,height=3.5in]{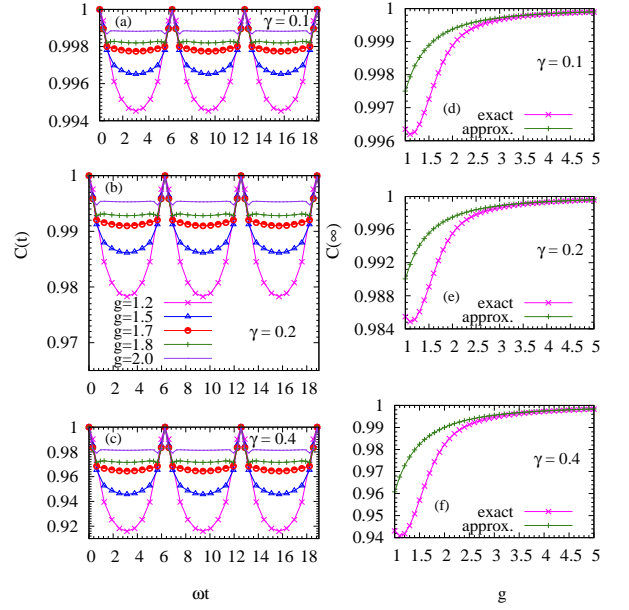}}
\caption{The figures (a), (b), and (c) in the left column represent the time variation of the {{coherence}} factor {C(t)} at various
values of $g$ and for three different values of 
$\gamma=\frac{J_\perp}{g\omega}$. In the right column, figures (d), (e), and (f) describe {${\rm C(\infty)}$} calculated 
approximately analytically
and 
{{ exactly}}. {{For large values of $g$ (e.g., $g=2$), these ${\rm C(\infty)}$ approximately match 
with the C(t) values in the nearly flat region}} between 
$2n \pi$ and $2 (n+1) \pi$ values of $\omega t$.}
\label{fig}
\end{figure}

 We will study the time variation, as well as the long-time behavior, of the matrix elements of the density matrix
by taking their initial values to be real. 
{{We define the {{coherence}} factor {C(t)}$\equiv \frac{|\langle \varepsilon_s | \rho_s(t)|\varepsilon_t\rangle|}
 {|\langle \varepsilon_s | \rho_s(0)|\varepsilon_t\rangle|}$ 
and the normalized difference of the diagonal elements (i.e., the inelastic factor) P(t)$\equiv
\frac{\langle \varepsilon_s | \rho_s(t)|\varepsilon_s\rangle-\langle \varepsilon_t | \rho_s(t)|\varepsilon_t\rangle}
{\langle \varepsilon_s | \rho_s(0)|\varepsilon_s\rangle - \langle \varepsilon_t | \rho_s(0)|\varepsilon_t\rangle}$; it
should be noted that ${\langle \varepsilon_s | \rho_s(t)|\varepsilon_s\rangle-\langle \varepsilon_t | \rho_s(t)|\varepsilon_t\rangle}$
also represents the population difference in the two $S^z_T=0$ states (i.e., the singlet and the $S^z_T=0$ triplet). 
We plot the  variation of C(t) and P(t)
 in  Figs. (\ref{fig}) and (\ref{fig1}), respectively, for different values of $g$ and $\gamma$.}}
 We see that, at $2n \pi$ (with $n=1,2,3,...$) values of $\omega t$, both C(t) and P(t) attain the initial values because all the harmonics
 ($\cos(\omega_{m_1, m_2}t)$)   present in the exponents of equations (\ref{sol_offdiag1}), (\ref{sol_offdiag2}), (\ref{sol_diag1}),
 and (\ref{sol_diag2}) are unity at this point. In other words, 
 all the harmonics are in phase at even multiples of $\pi$. But, at $\omega t=(2n+1) \pi$, 
 only the even (i.e.,  $m_1+m_2$ even) harmonics are unity whereas the odd harmonics are $-1$. 
Because of this, the probability difference P(t) sharply reaches the minimum at 
 $\omega t=(2n+1) \pi$ for all values of $g$. For off-diagonal elements, the region between two consecutive even $\pi$ values becomes flatter as 
 the $g$ value increases.
 At lower values of $g$, only a few lower harmonics (corresponding to a fewer number of phonons) contribute to this region because the 
 coefficients {$ C_{m}$} fall sharply as the harmonics become higher.
 On the other hand, with increasing values of $g$, more number 
 of harmonics contribute leading to a pronounced destructive interference.
 This results in the gradual flattening of the region with increasing $g$.
 {{Moreover, Fig. (\ref{fig}) 
 also tells us that, at sufficiently large values of $g$, 
the values 
  of {\rm C(t)} in the nearly flat region (at $\omega t$ away from $2n\pi$) match with their large time values {\rm C($\infty$)}.
 The same argument holds for the flattening of the 
{\rm  P(t)} (versus $\omega t$) curves between consecutive $\pi$ values as depicted in Fig. (\ref{fig1}).
{{As regards {\rm C($\infty$)} and {\rm P($\infty$)} curves in Figs. (\ref{fig}) and (\ref{fig1}), 
for obtaining our exact curves we  used a numerical approach;
whereas for our approximate analytic result,  
we used a large $g$ approximation as given in Appendix A.
 It is clear from the {\rm C($\infty$)} and {\rm P($\infty$)} }} figures that the coherence 
and the inelastic factor decrease 
 as the parameter 
 $\gamma=\frac{J_\perp}{g \omega}$ increases as can be seen from Eqs. (\ref{longtime}), (\ref{longtime1}), and (\ref{longtime3})}}.
{{Finally, we observe that C($\infty$) should correspond to the timescale at which the harmonics $\cos(\omega_m t)$ 
[occurring in equations (\ref{sol_offdiag1}), (\ref{sol_offdiag2}), (\ref{sol_diag1}),
 and (\ref{sol_diag2})]
oscillate many times 
to produce a vanishing contribution on an average. However, there is an additional oscillation (with time scale 
$ \hbar/J_{\perp}e^{-g^2} >> \hbar/\omega$) 
due to the presence of the system excitation gap [see Eq. (\ref{realsol}) in Appendix B]. Therefore, 
since the equilibrium time is much larger than the largest time scale, the coherence factor C(t) 
 attains a longtime value of {\rm C($\infty$)} at times $t \gg \hbar/(J_{\perp}e^{-g^2})$.}}
 Next, for a Markov process, we show that there won't be any decoherence for the local phonon case.
\begin{figure}
\centerline{\includegraphics[angle=90,angle=90,angle=90,width=3.2in,height=3.5in]{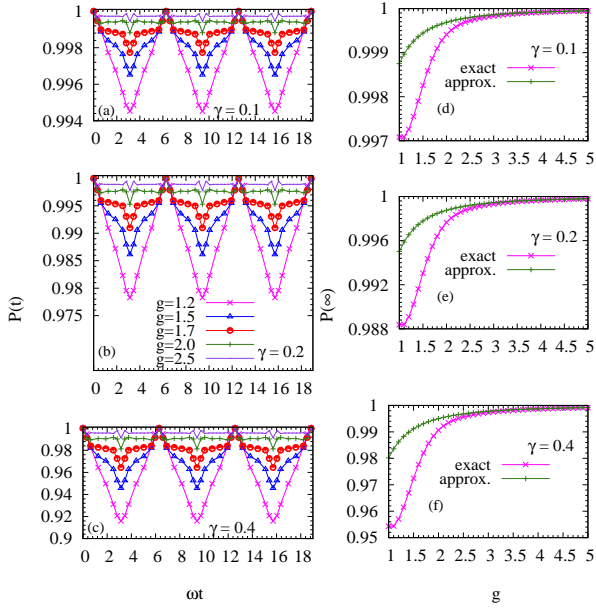}}
\caption{The figures (a), (b), and (c) in the left column represent the time variation of P(t) versus $\omega t$
plotted at various values of $g$ and for three different values of 
$\gamma=\frac{J_\perp}{g\omega}$. Figures (d), (e), and (f) (in the right column) describe ${\rm P(\infty)}$
 calculated approximately analytically 
and 
{{ exactly}}. {{ For large values of $g$ (e.g., $g=2.5$), these ${\rm P(\infty)}$ approximately match 
with the  P(t) values in the nearly flat region}} occurring at $\omega t$ values between two consecutive multiples of $\pi$.}
\label{fig1}
\end{figure}
\section{Markovian dynamics} 
Here (since $ J_\perp e^{-{g^2}} \ll \omega$), we assume that the relaxation time scale $\tau_s$ for the system  is much 
 larger than the correlation time $\tau_c$ for the environmental fluctuations, i.e., $\tau_s \gg \tau_c$.
In equation (\ref{mas2}), we replace $\tau$ by $(t-\tau)$; then, in the resulting equation,
 based on the assumption $\tau_s \gg \tau_c$, we replace the 
upper limit of the integration by $\infty$ \cite{open}. Thus, we arrive at the following Markovian master 
equation that is second-order in the perturbation $\tilde{H}_I^L$: 
\begin{eqnarray}
{ \frac{d \tilde{\rho}_s(t)}{dt}
= - \int_0^\infty d\tau Tr_B[\tilde{H}_I^L(t),[\tilde{H}_I^L(t-\tau), \tilde{\rho}_s(t) \otimes B_0]] .}\nonumber \\
\end{eqnarray}
By adopting the same analysis as in the previous section, we get
\begin{widetext}
\begin{eqnarray}
\frac{d \tilde{\rho}_s(t)}{dt} 
 &&= - \int_0^\infty d\tau \Big[\sum_{m} [ |_{ph}\langle 0 | H_I^L |m \rangle_{ph}|^2
 ~\tilde{\rho}_s(t)  e^{-i\omega_{m}\tau} 
\nonumber
 + \tilde{\rho}_s(t) ~ |_{ph}\langle 0 | H_I^L |m\rangle_{ph}|^2 
  e^{i\omega_{m} \tau}] \\
 &&~~~~~~~~~~~~~~~~ - \sum_{n}  [ _{ph}\langle n | H_I^L |0\rangle_{ph}  \tilde{\rho}_s(t) _{ph}\langle 0|
 H_I^L|n \rangle_{ph} e^{i\omega_{n}\tau}
+  {_{ph}}\langle n | H_I^L |0\rangle_{ph}  \tilde{\rho}_s(t) _{ph}\langle 0| H_I^L|n \rangle_{ph}
 e^{-i\omega_{n} \tau}] \Big] \nonumber \\
 &&= - \sum_{n}  
\left[ \int_0^{\infty} d\tau ~ e^{- i(\omega_{n} -i \eta)\tau} |_{ph}\langle 0|H_I^L |n \rangle_{ph}|^2 ~ \tilde{\rho}_s(t)
\right . 
+
\int_{0}^{\infty} d\tau ~ e^{i(\omega_{n}+i\eta)\tau} ~\tilde{\rho}_s(t) ~|_{ph}\langle 0 |H_I^L |n \rangle_{ph}|^2  \nonumber \\
&&~~~~~~~~~~~~ - \left. \int_{-\infty}^{\infty} d\tau ~ e^{i\omega_{n} \tau} ~ _{ph}\langle n| H_I^L |0 \rangle_{ph} 
~\tilde{\rho}_s(t)~ _{ph}\langle 0|H_I^L|n\rangle_{ph} 
    \right].
\end{eqnarray}
\end{widetext}
Now, using the relation $\int_{-\infty}^{\infty} d\tau e^{i \omega_{n} \tau} \propto \delta(\omega_{n})$ and equation (\ref{1order}), 
we end up getting 
\begin{eqnarray}
 &&\frac{d \tilde{\rho}_s(t)}{dt}\nonumber \\
 &&=  i~\sum_{n} \left[
  \frac{ | _{ph}\langle 0 |H_I^L |n \rangle_{ph}|^2 }{\omega_{n}} \tilde{\rho}_s(t) - 
\tilde{\rho}_s(t) \frac{ |_{ph}\langle 0 |H_I^L |n \rangle_{ph}|^2 }{\omega_{n}} 
\right]  .\nonumber \\
\label{markov}
\end{eqnarray}
{
The term $\sum_{n_1,n_2} \frac{ | _{ph}\langle 0 |H_I^L |n \rangle_{ph}|^2 }{\omega_{n}}$
appearing in the above equation, represents the second order correction term
in the effective Hamiltonian $H_{\rm eff}$ (obtained from second order perturbation theory);
 this term gives rise to the process in which both the spins simultaneously flip twice  to retain 
 the original spin configuration [or on invoking the equivalence between a spin and a hard-core-boson (HCB),
the process corresponds to the HCB particle hopping back and forth between the two sites].
From Eqs. (\ref{correction}) and (\ref{1order}), one obtains (as can be seen from Appendix A in Ref. \onlinecite{sahinur}) 
\begin{eqnarray}
H_{\rm eff} - H_s^{L} &=& -\sum_{m_1,m_2} \frac{ | _{ph}\langle 0 |H_I^L |n \rangle_{ph}|^2 }{\omega_{m}}
\nonumber \\
 &=&  \sum^{\prime}_{m_1,m_2} \frac{KC_m}{ \omega_m}
 (S^z_1S^z_2 -\frac{1}{4}) ,
\label{dom_2}
\end{eqnarray}
where for large values of $g^2$, $\sum^{\prime}_{m_1,m_2} \frac{KC_m}{ \omega_m} \approx \frac{J^2_\perp}{4 g^2 \omega}$ (see Ref. \onlinecite{sdadsy}).
Thus the total  
 correction term in the effective Hamiltonian
commutes with the $(S_{1}^{x}S_{2}^{x}+S_{1}^{y}S_{2}^{y})$ term. 
 Consequently, the eigenstates of the effective Hamiltonian remain the same as those of $H_s^{L}$. 
Therefore, by taking the matrix elements with respect to the $S_T^z=0$ states on both sides of the equation (\ref{markov}), we obtain the solutions }
\begin{eqnarray}
 \langle \varepsilon_s | \rho_s(t)|\varepsilon_t\rangle&=& \langle \varepsilon_s | \rho_s(0)|\varepsilon_t\rangle e^{-i(\varepsilon_s-\varepsilon_t)t}, \\
 \langle \varepsilon_t | \rho_s(t)|\varepsilon_s\rangle&=& \langle \varepsilon_t | \rho_s(0)|\varepsilon_s\rangle e^{-i(\varepsilon_t-\varepsilon_s)t},\\
 \langle \varepsilon_s | \rho_s(t)|\varepsilon_s\rangle&=& \langle \varepsilon_s | \rho_s(0)|\varepsilon_s\rangle, 
\end{eqnarray}
 and 
\begin{eqnarray}
 \langle \varepsilon_t | \rho_s(t)|\varepsilon_t\rangle&=& \langle \varepsilon_t | \rho_s(0)|\varepsilon_t\rangle.
\end{eqnarray}
Therefore, for Markovian dynamics no decoherence is observed up to second order in perturbation!
\section{Global phonon case}
Here we consider the model where both the spins are identically coupled to the same set of phonons
 (i.e., spins are globally coupled to a bath). The total Hamiltonian is given by
\begin{eqnarray}
\label{globalham}
 H&=&H_s + H_{B} + H_I = H_s +   \omega a^{\dagger} a+  g \omega S^z_{\rm Total}  (a^{\dagger} + a),\nonumber \\
\end{eqnarray}
where $S_{\rm Total}^z=\sum_{i=1,2} S^z_i$. Since $S^z_{\rm Total}$ commutes with $H_s$, the eigenstates of $H_s$ having the same eigenvalues $S_{T}^z$ 
 form a DFS. But we also would like to have the form of $\rho_s^O(t)$ (the density matrix for the system in the original lab frame) so that we can 
analyze the case when $S_{T}^z$ is not conserved. We start with the non-Markovian 
master equation \cite{yu} given by 
\begin{equation}
 \frac{d \rho_s^O(t)}{dt}
=-i[H_s,\rho_s^O(t)] + F(t)[L\rho_s^O(t),L] 
+F^{\star}(t)[L,\rho_s^O(t) L] ,
\label{master}
\end{equation}
where L is the system operator coupled to the environment (i.e., $L=S^z_{\rm Total}$) and  satisfies the condition $[L,H_s]=0$. 
In the above equation, $ F(t)=\int_0^{t} \alpha(t-s)ds$ where the environmental correlation function (at temperature T)
 is given by $\alpha(t-s)= \eta(t-s)+i\nu(t-s)$ with 
\begin{eqnarray}
\eta(t-s) = g^2 \omega^2\coth(\frac{\omega}{2k_B T} )\cos[\omega(t-s)] ,  
\end{eqnarray}
{and}  
\begin{eqnarray}
\nu(t-s) =- g^2 \omega^2 \sin[\omega(t-s)] .
\end{eqnarray}
The above non-Markovian master equation (\ref{master}) is valid at all coupling strengths and when the system and the environment are initially uncorrelated.
The complete set of the system eigenstates $\{|\varepsilon\rangle_s\}$ are the simultaneous eigenstates of $S_{\rm Total}^z$ and $H_s$.
 Thus, on  solving the master equation 
(\ref{master}), we get
\begin{eqnarray}
 &&{_s\langle \varepsilon|\rho_s^O (t)|\varepsilon^{\prime}\rangle_s} \nonumber \\
    &&~~=  \exp \left (-i \left [(\varepsilon -\varepsilon^{\prime})t + 
\left \{ (S^z_{T\varepsilon})^2-(S^z_{T\varepsilon^{\prime}})^2 \right \} Y(t) \right ] \right )
\nonumber \\  
 &&~~~~~~~\times   \exp \left [- \left ( S^z_{T\varepsilon}-S^z_{T\varepsilon^{\prime}} \right )^2 X(t) \right ]
~{_s\langle\varepsilon |\rho_s^O (0)|\varepsilon^{\prime}\rangle_s} ,\nonumber \\
\label{globalsol}
\end{eqnarray}
where $\varepsilon$ and $S^z_{T\varepsilon}$ are recognized from $H_{s}|\varepsilon\rangle_s = \varepsilon|\varepsilon\rangle_s$ and $S_{\rm Total}^z
|\varepsilon \rangle_s =S^z_{T\varepsilon}|\varepsilon\rangle_s$. Moreover, $X(t) \equiv \int_0^{t} F_R(s)ds= g^2 \coth(\frac{\omega}{2k_B T} )
[1-\cos(\omega t)]$ and $Y(t) \equiv \int_0^{t} F_I(s)ds=g^2[- \omega t + \sin(\omega t)] $ 
with $F_R(t) + i F_I(t) \equiv F(t)$. The solution given by Eq. (\ref{globalsol}) shows decoherence free behavior for the states $\{|\varepsilon\rangle_s \}$ 
having the same $S_{T}^z$ values. But when $S_{T}^z$ is not conserved (i.e., $S^z_{T\varepsilon} \neq S^z_{T\varepsilon^{\prime}}$), decoherence will occur 
due to the factor $\exp \left [- \left ( S^z_{T\varepsilon}-S^z_{T\varepsilon^{\prime}} \right )^2 g^2 \coth(\frac{\omega}{2k_B T} )
[1-\cos(\omega t)] \right ]$.
{{We give below, explicit matrix elements with respect to the $S_T^z=0$ states
\begin{eqnarray}
 \langle \varepsilon_s | \rho_s^O(t)|\varepsilon_t\rangle&=& \langle \varepsilon_s | \rho_s^O(0)|\varepsilon_t\rangle e^{-i(\varepsilon_s-\varepsilon_t)t},
\end{eqnarray}
 and 
\begin{eqnarray}
 \langle \varepsilon_t | \rho_s^O(t)|\varepsilon_t\rangle&=& \langle \varepsilon_t | \rho_s^O(0)|\varepsilon_t\rangle.
\end{eqnarray}
}}

{{Finally, it is worth noting that the case of global noise in the Lang-Firsov frame of reference is a trivial
problem as the transformed Hamiltonian $H^L$ is uncoupled; consequently, a simply separable initial state $\rho_s (0) \otimes B_0$
leads to a simply separable state $\rho_s (t) \otimes B_0$ at all times!}} 
{{
\section{
Bang-bang control}
To achieve dynamical decoupling, we use a series of instantaneous pi pulses that flip both the spins simultaneously. 
Now, in the $S_T^z=0$ subspace,
the operator for these pulses is given by 
$P_{\pi}= S_{1}^{+}S_{2}^{-}+S_{2}^{+}S_{1}^{-}$; the pulses are applied to the system with an interval $\delta t$ which is small 
enough so that the error in the estimation of decoherence
can be restricted to $\textit{O}(\delta t)$. Next, we calculate the composite evolution operator 
$U_1(t,\delta t)$ corresponding to the evolution from time $t$ to time $t+ 2\delta t$; it is given by 
$U_1(t,\delta t)=\tilde{U}(t+2\delta t,t+\delta t) P_{\pi} \tilde{U}(t+\delta t,t) P_{\pi}$ where $\tilde{U}(t,t^{\prime})$ is the time evolution operator for the 
total system in the interaction picture and  is given by
\begin{eqnarray}
\label{operator}
 \tilde{U}(t,t^{'}) &&=\tilde{U}(t,0)\tilde{U}^{\dagger}(t^{'},0)\nonumber \\
 &&=e^{iH_{0}t} e^{-iH t} e^{iH t^{'}} e^{-iH_{0} t^{'}}\nonumber \\
 &&=e^{iH_{0}t} e^{-iH(t-t^{'})}  e^{-iH_{0}t^{'}}.
\end{eqnarray}
Therefore,
\begin{eqnarray}
\label{compositeop}
 U_1(t,\delta t) &&=\tilde{U}(t+2\delta t,t+\delta t) P_{\pi} \tilde{U}(t+\delta t,t) P_{\pi} \nonumber \\
 &&= e^{iH_{0}(t+2\delta t)} e^{-iH \delta t}  e^{-iH_{0}(t+\delta t)} P_{\pi}
\nonumber \\
&&~~~~~\times e^{iH_{0}(t+\delta t)} e^{-iH \delta t}  e^{-iH_{0}t} P_{\pi} \nonumber\\
 &&= e^{iH_{0}(t+2\delta t)} e^{-iH \delta t}P_{\pi}
 e^{-iH \delta t}P_{\pi}e^{-iH_{0}t},
\end{eqnarray}
{{where, use has been made of the fact that $[H_0,P_{\pi}]=0$. Now, neglecting the $\textit{O}(\delta t^2)$ term $[H_{0},H_{I}]\delta t^2$ and higher
order terms in $\delta t$, we write
\begin{eqnarray}
 e^{-iH \delta t} \approx e^{-iH_{0} \delta t}e^{-iH_{I} \delta t} .
\label{odt2}
\end{eqnarray}
 Furthermore, we note that, {{since $H_I= g\omega\sum_{i=1,2} S_{i}^{z}( a_i + a_{i}^{\dagger})$,}}
\begin{equation}
 P_{\pi} e^{-i(H_0+H_I) \delta t}=e^{-i(H_0-H_I) \delta t}P_{\pi},
\label{pi}
 \end{equation}
which means that the pulse changes the sign of the interaction term.
Therefore, using the above two Eqs. (\ref{odt2}) and (\ref{pi}), we simplify equation (\ref{compositeop}) as}}
\begin{eqnarray}
U_1(t, \delta t)=\tilde{U}(t+2\delta t,t+\delta t) P_{\pi} \tilde{U}(t+\delta t,t) P_{\pi} 
=  I + \textit{O}(\delta t^2).\nonumber \\
  \label{evolution}
\end{eqnarray}
 Now, in the interaction picture $\tilde{\rho}_T^O$ (the total density matrix in the original lab frame) is given by
\begin{eqnarray}
 \tilde{\rho}_T^O (2\delta t)&=&\tilde{U}(2\delta t,\delta t) P_{\pi} \tilde{U}(\delta t,0) P_{\pi} \tilde{\rho}_T^O (0) 
\nonumber \\
&&~~~\times  P_{\pi} \tilde{U}^{\dagger}(\delta t,0) P_{\pi} \tilde{U}^{\dagger}(2\delta t,\delta t)
\nonumber \\
&=& \rho_{T}^O(0) + \textit{O}(\delta t^2).
\end{eqnarray}
Hence, in the interaction picture
we obtain $\tilde{\rho}_s^O(t)$ (the density matrix for the system in the original lab frame) to be
\begin{eqnarray}
 \tilde{\rho}_s^O (2\delta t) &&= 
{\sum_{n_1,n_2} {_{ph}\langle n_1,n_2 |}\tilde{\rho}_T^O (2\delta t)| n_1,n_2\rangle_{ph}} \nonumber \\
&&=
\sum_{n_1,n_2}~ _{ph}\langle n_1,n_2 | \rho_s^O (0) \otimes B_0 | n_1,n_2\rangle_{ph}
 + \textit{O}(\delta t^2) \nonumber \\
  &&=\rho_s^O(0)+\textit{O}(\delta t^2).
\end{eqnarray}
 If we make 
the interval  between successive pulses $\delta t \rightarrow 0$ (i.e., $\delta t$ is much smaller than the time taken by
 the environment to produce decoherence), 
the evolution of 
the density matrix is nearly unitary, i.e., $\rho_s^O(2\delta t) \approx  e^{-iH_{s}2\delta t} \rho_s^O(0)  e^{iH_{s}2\delta t}$. In general, if 
$\delta t= \frac{t}{N}$ 
\begin{eqnarray}
 \rho_{T}^O(t) &=& e^{-iH_0 t}\Big[ I + \textit{O}(\delta t^2)\Big]^{\frac{N}{2}} \rho_{T}^O(0)  \Big[I+ \textit{O}(\delta t^2)\Big]^{\frac{N}{2}}e^{iH_0 t} .
 \nonumber \\
\end{eqnarray}
Hence, for error up to $\textit{O}(\delta t)$, we get
\begin{eqnarray}
\lim_{N\rightarrow \infty} \rho_{T}^O(t) &=& e^{-iH_{0}t } \rho_{T}^O(0)  e^{iH_{0}t} .
\end{eqnarray}
It then follows that
\begin{eqnarray}
 \rho_s^O (t)&=& e^{-iH_{s}t } \rho_{s}^O(0)  e^{iH_{s}t}.
 \label{ncycle}
 \end{eqnarray}
Now, the evolution operator $e^{-iH \delta t}$ can be written as 
 $ e^{-i(H_0+H_I) \delta t} \approx e^{-iH_{0} \delta t}e^{-iH_{I} \delta t}e^{\frac{1}{2}[H_0,H_I]\delta t^2}+O(\delta t^3)$.
 Dynamical decoupling up to next order in $\delta t$ [as compared to $U_1(t,\delta t)$ in Eq. (\ref{evolution})]
can be obtained by using the unequally spaced pulse sequence in the composite operator given below: 
  \begin{eqnarray}
 \nonumber U_2(t,\delta t)&=&U_1(t+2\delta t, \delta t) P_{\pi}U_1(t, \delta t)P_{\pi} \\
 \nonumber &=&\tilde{U}(t+4\delta t,t+3\delta t) P_{\pi} \tilde{U}(t+3\delta t,t+2\delta t)\nonumber \\
 &&~~\times\tilde{U}(t+2\delta t,t+\delta t) P_{\pi} \tilde{U}(t+\delta t,t) \nonumber \\
 &=&I+\textit{O}[\delta t^3] ,
\label{U2}
 \end{eqnarray}
where $U_2(t,\delta t)$ corresponds to the evolution from time $t$ to time $t+ 4\delta t$.
 Here, we should mention that higher order decoupling (than that achieved above using $U_2$)
cannot be realized in this case as the commutators involving even number of $H_{I}$
terms do not change sign 
 upon the application of $P_{\pi}$ pulses.}}

\begin{widetext}
{{
We will now explicitly evaluate the time dependence of $U_2(t,\delta t)$. Using Eq. (\ref{pi}), we get the following expression for $U_2$:
\begin{eqnarray}
 U_2(t,\delta t)= 
e^{iH_{0}(t+4\delta t)}
\Bigg( e^{-iH_+ \delta t} e^{-iH_- \delta t} \Bigg)
 \Bigg[ e^{-iH_- \delta t}e^{-iH_+ \delta t} \Bigg ]
e^{-iH_{0}t} ,
\label{U2_2}
 \end{eqnarray}
where $H_{\pm}=H_0 \pm H_I$ (with  $H_+ \equiv H$).
Then using the following Zassenhaus formula \cite{casas} (which is related to the well-known  Campbell-Baker-Hausdorff formula)
\begin{eqnarray}
\!\!\! e^{t(X+Y)}
=e^{tX}e^{tY}e^{-\frac{t^2}{2}[X,Y]}
e^{\frac{t^3}{6}\left(2[Y,[X,Y]]+[X,[X,Y]]\right )}
 e^{-\frac{t^4}{24}\left([[[X,Y],X],X]+3[[[X,Y],X],Y]+3[[[X,Y],Y],Y]\right )}..... ,
\end{eqnarray}
 to the terms $\Big( e^{-iH_+ \delta t} e^{-iH_- \delta t} \Big)$ and
$\Big[ e^{-iH_- \delta t}e^{-iH_+ \delta t} \Big]$
 in the expression for $U_2$ in Eq. (\ref{U2_2}), we get (for error up to $O(\delta t^5)$) 
\begin{eqnarray}
&&
 U_2(t,\delta t)
=
e^{iH_{0}(t+4\delta t)}
\nonumber \\
&& \times
 \Bigg( e^{-i2 H_0\delta t}  
e^{\frac{\delta t^4}{24}\left([[[H_+,H_-],H_+],H_+]+3[[[H_+,H_-],H_+],H_-]+3[[[H_+,H_-],H_-],H_-]\right )}
  e^{\frac{-i\delta t^3}{6}\left(2[H_-,[H_+,H_-]]+[H_+,[H_+,H_-]]\right )}
e^{-\frac{\delta t^2}{2}[H_+,H_-]} \Bigg)
\nonumber \\
&& \times \Bigg[ e^{\frac{\delta t^2}{2}[H_+,H_-]}
e^{\frac{-i\delta t^3}{6}\left(2[H_-,[H_+,H_-]]+[H_+,[H_+,H_-]]\right )} 
  e^{\frac{-\delta t^4}{24}\left([[[H_+,H_-],H_+],H_+]+3[[[H_+,H_-],H_+],H_-]+3[[[H_+,H_-],H_-],H_-]\right )}
  e^{-i2 H_0\delta t} \Bigg]
\nonumber \\
&& {{\times e^{-iH_{0}t}}}
\nonumber \\
&&=
e^{iH_{0}(t+4\delta t)}
 e^{-i2 H_0\delta t}  
  e^{\frac{-i2 \delta t^3}{6}\left(2[H_-,[H_+,H_-]]+[H_+,[H_+,H_-]]\right )}
  e^{-i2 H_0\delta t}
 e^{-iH_{0}t}   . 
\label{U2_3}
 \end{eqnarray}
}}

{{
Now, the total evolution operator (characterizing evolution starting at zero time and ending at time t) is given by
\begin{eqnarray}
\!\!\! U_2^T(t) 
 = 
U_2(t-4\delta t,\delta t)U_2(t-8\delta t,\delta t).....U_2(4\delta t,\delta t)U_2(0,\delta t)
 = e^{iH_{0}t}
 \Bigg [e^{-i2 H_0\delta t}  
 \left (1+ \delta t^3 A \right )
  e^{-i2 H_0\delta t} +\textit{O}[\delta t^5]\Bigg ]^{\frac{N}{4}}  ,
\label{U2_4} 
\end{eqnarray}
where $A\equiv \frac{-i}{3}\left(2[H_-,[H_+,H_-]]+[H_+,[H_+,H_-]]\right )$. Then the density matrix, in 
the interaction picture, is expressed as
\begin{eqnarray}
 \tilde{\rho}_T^O (t)={U}_2^T(t) \tilde{\rho}_T^O (0)  {U}_2^{T\dagger}(t) ,
\end{eqnarray}
which in the Schr\"odinger representation is given by 
\begin{eqnarray}
 {\rho}_T^O (t)=e^{-iH_0t}{U}_2^T(t) \left [ {\rho}_s^O (0)\otimes B_0 \right ] {U}_2^{T\dagger}(t)e^{iH_0t} .
\end{eqnarray}
 Based on Eq. (\ref{U2_4}), we note that $U_2^T(t)=1+O(\delta t^2)$; consequently we get the expression 
\begin{eqnarray}
 {\rho}_T^O (t)=e^{-iH_0t}\Big (
 {U}_2^T(t) \left [ {\rho}_s^O (0)\otimes B_0 \right ]  
+ \left [ {\rho}_s^O (0)\otimes B_0 \right ]{U}_2^{T\dagger}(t)
-\left [ {\rho}_s^O (0)\otimes B_0 \right ]\Big ) e^{iH_0t} 
+\textit{O}[\delta t^4] .
\end{eqnarray}
Hence, it follows that 
\begin{eqnarray}
\rho_s^O(t)
&&
 =\sum_n {_{ph}}
\langle n|{\rho}_T^O (t)|n\rangle_{ph}
\nonumber \\
&&
 =e^{-iH_st}
\Bigg [ \sum_n \Big ( {_{ph}}\langle n|{U}_2^T(t)|n\rangle_{ph} ~ {\rho}_s^O (0)  
+  {\rho}_s^O (0)\sum_n {_{ph}}\langle n|{U}_2^{T\dagger}(t)|n\rangle_{ph}
 - {\rho}_s^O (0)\Big )\frac{e^{-\beta\omega_n}}{Z} \Bigg ]e^{iH_st}  .
\label{rhost}
\end{eqnarray}
Next, it is important to realize that
\begin{eqnarray}
{_{ph}}\langle n|\Big(2[H_-,[H_+,H_-]]+[H_+,[H_+,H_-]]\Big ) |n\rangle_{ph}
&&
={_{ph}}\langle n|\Big(6[H_0,[H_I,H_0]]-2[H_I,[H_I,H_0]]\Big ) |n\rangle_{ph}
\nonumber \\
&&
={_{ph}}\langle n|\Big(-2[H_I,[H_I,H_0]]\Big ) |n\rangle_{ph} ,
\label{Hcom}
\end{eqnarray}
as only even powers of $H_I$ survive the expectation value in the state $|n\rangle_{ph}$.
Then, on noting that $H_I= g\omega\sum_{i=1,2} S_{i}^{z}( a_i + a_{i}^{\dagger})$ and that $H_0
=J_\lVert S_{1}^{z}S_{2}^{z} - J_\perp (S_{1}^{x}S_{2}^{x}+S_{1}^{y}S_{2}^{y}) 
+ \omega \sum_{i=1,2} a_{i}^{\dagger} a_i$, we get the following relevant expression
\begin{eqnarray}
[H_I,[H_I,H_0]]
= 
 - J_\perp g^2 \omega^2 \Big (S_{1}^{x}S_{2}^{x}+S_{1}^{y}S_{2}^{y} \Big )\Big [(a_2+a_2^{\dagger})-(a_1+a_1^{\dagger}) \Big]^2
 -2g^2\omega^3 \Big (S_1^{z2}+S_{2}^{z2} \Big ) .
\label{HI}
\end{eqnarray}
To evaluate $\langle \varepsilon|{_{ph}}\langle n|U_2^T(t)|n\rangle_{ph}$, we use Eqs. (\ref{U2_4}), (\ref{Hcom}), and (\ref{HI}) to obtain
\begin{eqnarray}
\langle \varepsilon|~{_{ph}}\!\langle n|\left ( 1+A\delta t^3 \right )|n\rangle_{ph}
&&=\langle \varepsilon|\Big (1-i\frac{2}{3}g^2\omega^3\delta t^3 \Big )
 -i\frac{\alpha_{\varepsilon}}{3}J_{\perp}g^2\omega^2 \delta t^3 
\langle \varepsilon|~ {_{ph}}\!\langle n|
\Big [(a_2+a_2^{\dagger})-(a_1+a_1^{\dagger}) \Big]^2
|n\rangle_{ph} 
\nonumber \\
&&=\langle \varepsilon|\Big (1-i\frac{2}{3}g^2\omega^3\delta t^3 
-i\frac{2}{3}\alpha_{\varepsilon}J_{\perp}g^2\omega^2 \delta t^3 (n+1) \Big )  ,
\end{eqnarray}
where $n=n_1+n_2$ and $\alpha_{\varepsilon}=-1$ $(1)$ when $ |\varepsilon \rangle$ represents a singlet ($S^z_T=0$ triplet) state;
it then follows that 
\begin{eqnarray}
\langle \varepsilon|~{_{ph}}\!\langle n|U_2^T(t)|n\rangle_{ph}
=\langle \varepsilon|\Big (1-i\frac{2}{3}\frac{N}{4}g^2\omega^3\delta t^3 
-i\frac{2}{3}\frac{N}{4}\alpha_{\varepsilon}J_{\perp}g^2\omega^2 \delta t^3 (n+1) \Big ) +\textit{O}[\delta t^4] .
\label{U2_matrix}
\end{eqnarray}
From Eqs. (\ref{rhost}) and (\ref{U2_matrix}) and the relation $\delta t N =t$, for error up to $O(\delta t^4)$, we finally obtain 
\begin{eqnarray}
\langle \varepsilon|\rho_s^O(t)| \varepsilon^{\prime}\rangle =
\Bigg [1 
-i\sum_n \Big (
 (\alpha_{\varepsilon}-\alpha_{\varepsilon^{\prime}}) J_{\perp}g^2\omega^2 (n+1) \frac{t \delta t^2}{6} \Big )
\frac{e^{-\beta\omega_n}}{Z} \Bigg ]\langle \varepsilon|\rho_s^O(0)| \varepsilon^{\prime}\rangle e^{-i(\varepsilon-\varepsilon^{\prime})t} .
\end{eqnarray}
At ${\rm T = 0~K}$, we get the simpler expression 
\begin{eqnarray}
\langle \varepsilon|\rho_s^O(t)| \varepsilon^{\prime}\rangle =
 \langle \varepsilon|\rho_s^O(0)| \varepsilon^{\prime}\rangle  e^{-i(\varepsilon-\varepsilon^{\prime})t}
e^{ 
-\left (\frac{i}{6}  (\alpha_{\varepsilon}-\alpha_{\varepsilon^{\prime}}) J_{\perp}g^2\omega^2  t \delta t^2 \right )} ,
\end{eqnarray}
which implies the system is free of decoherence with order of magnitude of the error being at most  $\delta t^4$.}}
{{We will now give explicit expressions for the matrix elements with respect to the singlet and $S^z_T=0$ triplet states
\begin{eqnarray}
 \langle \varepsilon_s | \rho_s^O(t)|\varepsilon_t\rangle&=& \langle \varepsilon_s | \rho_s^O(0)|\varepsilon_t\rangle e^{-i(\varepsilon_s-\varepsilon_t)t}
~e^{i 
\big (\frac{J_{\perp}g^2\omega^2  \delta t^2 }{3}\big ) t} ,
\label{dt3}
\end{eqnarray}
 and 
\begin{eqnarray}
 \langle \varepsilon_t | \rho_s^O(t)|\varepsilon_t\rangle&=& \langle \varepsilon_t | \rho_s^O(0)|\varepsilon_t\rangle.
\end{eqnarray}
}}
{{In the above Eq. (\ref{dt3}), it should be noted that the off-diagonal matrix element involves only a phase factor.}}
\end{widetext}
\section{Tunnel-coupled double quantum dot system}
{{
Lastly, we propose an oxide (i.e., manganite) based double quantum dot (DQD) system with only one $e_g$ electron (capable of tunneling
between the dots) as an equivalent of a qubit. The logical qubit is formed from the two possible electron occupation states
 with $|0\rangle_{Q} \equiv |01 \rangle$ and 
$|1\rangle_{Q} \equiv |10\rangle$. 
The advantages of an oxide double-quantum-dot system are as follows:
(a) similar to semiconductor double quantum dots \cite{petta,Barthel}, here too fast electrical control of exchange interaction is possible;
(b) compared to semiconductors, the extent of the electronic wave function in oxides (which is about a lattice constant) is much smaller
and thus the size of the oxide quantum dot can be much smaller leading to being much better suited for miniaturization;
and (c) the decoherence due to optical phonons (the main source of noise)
in oxide dots is significantly smaller than the decoherence due to nuclear spins in
semiconductor dots. 
}}

{{
The dot with the $e_g$ electron can be treated as an up spin while
the dot without the $e_g$ electron can be regarded as a down spin. The tunneling of the $e_g$ electron between the dots and 
the attraction between
the electron and the hole on adjacent dots can be modelled as an anisotropic Heisenberg interaction between
two spins with the total z-component of the spins being zero. 
Tunneling between the dots can be controlled by a gate voltage.
}}

{{
We will now elaborate on the  preparation of a simply separable initial state $\rho_T(0)={\rho}_s(0) \otimes B_0$.
We start with the 
 gate voltage set to prohibit interdot tunneling (i.e., $J_{\perp}=0$); next, we introduce an
 electron in one of the quantum dots
to obtain the state $|10\rangle \otimes |0\rangle_{ph} = \frac{1}{\sqrt{2}}[|\varepsilon_s\rangle + |\varepsilon_t\rangle] \otimes |0\rangle_{ph}$
[as can be seen from equation (\ref{HLF})].
{{Here, it should be noted that 
the $|0\rangle_{ph}$ state is the shifted ground state (in the polaronic frame).}}
Then, we
introduce a small tunneling $J_{\perp}/\omega \ll 1$ (i.e., about $10^{-3}$ when $\hbar \omega \sim 0.05$ eV) 
by changing the gate voltage rapidly (in a period much smaller than $\hbar/J_{\perp} \sim 10$ ps 
so that {{sudden approximation}} is valid)
 and let the system evolve. 
{{As $\frac{J_\perp}{\omega}\ll1$, the phonon distortion quickly follows the location of the electron.}}
For a small value of $J_{\perp}/\omega$, 
 $|\varepsilon_s\rangle\otimes |0\rangle_{ph}$ and $ |\varepsilon_t\rangle\otimes |0\rangle_{ph}$ are approximate
eigenstates [of the Hamiltonian in the Lang-Firsov frame given by Eq. (\ref{HLF})]
with a probability larger than 
$1-J_{\perp}^2/(4 g^4 \omega^2)$ (i.e., greater than $ 0.999999$) for $g^2 \gg1$.
This is evident from the fact that, under the perturbation 
given by $H_I^L$ [in Eq. (\ref{hil})], the (first-order) correction terms to the states
$|\varepsilon_{s(t)}\rangle\otimes |0\rangle_{ph}$ are given by 
\begin{eqnarray}
 [|\varepsilon_{s}\rangle\langle \varepsilon_{s}|
+|\varepsilon_{t}\rangle\langle \varepsilon_{t}|]\otimes 
\sum_{n\neq 0}\frac{ |n\rangle_{ph}~{_{ph}}\langle n|H_I^L |\varepsilon_{s(t)}\rangle \otimes |0\rangle_{ph}}{-\omega_n} ;
\nonumber \\
\end{eqnarray}
thus the sum of the probabilities of all the correction terms is given by 
\begin{eqnarray}
 \frac{J_{\perp}^2}{4}\sum_n\frac{|\langle n|_{ph} e^{g(a_2-a^{\dagger}_2)}e^{-g(a_1-a^{\dagger}_1)} -e^{-g^2}|0\rangle_{ph}|^2}{\omega_n^2}
\approx \frac{J_{\perp}^2}{4 g^4 \omega^2} ,
\nonumber \\
 \end{eqnarray}
[see Ref. \onlinecite{sdadsy} for similar results].
Thus, the  evolved state [for $J_{\perp}/(g^2 \omega) \ll 1$] is a simply separable initial state
(in the {dressed basis})  given by:
\begin{eqnarray}
|\psi(t)\rangle = && \left (e^{-iH^Lt} \right ) \frac{1}{\sqrt{2}}[|\varepsilon_s\rangle + |\varepsilon_t\rangle] \otimes |0\rangle_{ph}
\nonumber \\
\approx && \frac{1}{\sqrt{2}}[(e^{-i\varepsilon_s t}|\varepsilon_s\rangle +(e^{-i\varepsilon_t t} |\varepsilon_t\rangle] \otimes |0\rangle_{ph}
\nonumber \\
\approx && \left [\cos \left (\frac{J_{\perp}e^{-g^2}t}{2} \right ) |10\rangle
+i\sin \left (\frac{J_{\perp}e^{-g^2}t}{2} \right )
 |01\rangle \right ]
\nonumber \\
&& \otimes |0\rangle_{ph}  .
\label{ev_st}
\end{eqnarray}
{{In the above approximation, the ignored (first-order) correction terms are given by
\begin{eqnarray}
 [|\varepsilon_{s}\rangle\langle \varepsilon_{s}|
+|\varepsilon_{t}\rangle\langle \varepsilon_{t}|]\otimes 
\sum_{n\neq 0}[1- e^{-i \omega_n t}]\frac{ |n\rangle_{ph}~{_{ph}}\langle n|H_I^L |\psi(0) \rangle}{-\omega_n} ,
\nonumber \\
\end{eqnarray}
with the sum of probabilities of all the correction terms being $\sim \frac{J_{\perp}^2}{ g^4 \omega^2}$.}}
{{Furthermore, it is also of interest to note that
the above equation (\ref{ev_st}) is consistent with Eqs. (\ref{sol_offdiag1}), (\ref{sol_offdiag2}), (\ref{sol_diag1}), and (\ref{sol_diag2}).}}
Upon introduction of a magnetic flux,
 the tunneling term 
$J_{\perp}$
 acquires an Aharnov-Bohm phase factor $e^{i\phi}$  and we obtain the
following general simply separable initial state:
\begin{eqnarray}
|\psi(t)\rangle  
= && \left [\cos \left (\frac{J_{\perp}e^{-g^2}t}{2} \right ) |10\rangle
+ie^{-i\phi}\sin \left (\frac{J_{\perp}e^{-g^2}t}{2} \right )
 |01\rangle \right ]
\nonumber \\
&& \otimes |0\rangle_{ph}  .
\end{eqnarray}
Lastly (after the desired simply separable initial state is achieved),
 by altering the gate voltage and the magnetic flux rapidly (i.e., much quicker than the tunneling time for the electron)
to get the desired value of the tunneling $J_{\perp}$,
the dynamics of the polarons and their decoherence
are studied.}}

{{
The readout of the evolved polaronic state (in the dressed basis)
can be obtained as follows. Let the evolved state in the Lang-Firsov
frame be given by $|\psi\rangle^L \approx [\alpha |01\rangle + \sqrt{1-|\alpha|^2} |10\rangle]\otimes |0\rangle_{ph}$ 
 Then the same state in the original lab frame of reference is given by $|\psi\rangle^O = e^{-S}|\psi\rangle^L$.
Then, using Eq. (\ref{HLF}), we obtain the following expression:
\begin{eqnarray}
 |\psi\rangle^O = [\alpha |01\rangle e^{-g(a_2-a^{\dagger}_2)}+\sqrt{1-|\alpha|^2} |10\rangle e^{-g(a_1-a^{\dagger}_1)}]\otimes|0\rangle_{ph} .
\nonumber \\
\end{eqnarray}
Then, the reduced density matrix $\rho_s^O$ in the original lab frame [obtained from total density matrix $\rho_T^O$ (in the lab frame)]
can be deduced as follows
\begin{eqnarray}
&& \langle 10 | \rho_s^O(t)|01\rangle
\nonumber \\
&& =\sum_n {_{ph}}\langle n| \langle10| \rho_T^O(t) |01\rangle |n\rangle_{ph} 
\nonumber \\
&& = {{\alpha^{\star}}} \sqrt{1-|\alpha|^2} \sum_n {_{ph}}\langle n|
e^{-g(a_1-a^{\dagger}_1)}|0\rangle_{ph}~{_{ph}}\langle 0| e^{g(a_2-a^{\dagger}_2)} |n \rangle_{ph}
\nonumber \\
&&={{\alpha^{\star}}} \sqrt{1-|\alpha|^2} e^{-g^2}
\nonumber \\
&& = e^{-g^2} \langle10| \rho_s(t)|01\rangle.
\end{eqnarray}
Thus, at ${\rm T= 0~  K}$, the off-diagonal elements of the system density matrix in the two frames of reference
differ only by a factor $e^{-g^2}$. Hence, one can readout (non-invasively) the off-diagonal elements
in the original frame of reference and deduce the 
result for the polaronic frame of reference. One can readout the charge state non-invasively by
various methods such as using 
a radio frequency resonant circuit coupled to a DQD as demonstrated in Ref. \onlinecite{ritchie};
employing quantum point contact charge detector as was done in Ref. \onlinecite{petta2};
transferring quantum information to a quantized cavity field as explained in Ref. \onlinecite{tan}.
}}

{{
Next, at 0 {\rm K}, we also give an alternate derivation of the result 
$\langle 10 | \rho_s^O(t)|01\rangle= e^{-g^2} \langle10| \rho_s(t)|01\rangle$.
 We first show the connection between
the matrix elements of the density matrix in the lab frame and the polaronic frame.
From the definition of the reduced density matrix $\rho_s^O$ in the original lab frame in terms of the total density
matrix $\rho^O_T$ in the original frame and the total density operator $\rho_T$ in the polaronic frame, we have
\begin{eqnarray}
 \langle \varepsilon | \rho_s^O(t)|\varepsilon^{\prime}\rangle
&&\equiv \langle \varepsilon | \sum_n {_{ph}}\langle n|\rho_T^O(t)||n \rangle_{ph}|\varepsilon^{\prime}\rangle
\nonumber \\
&&\equiv \langle \varepsilon | \sum_n {_{ph}}\langle n|e^{-S}\rho_T(t)e^S|n \rangle_{ph}|\varepsilon^{\prime}\rangle .
\end{eqnarray}
Then, since the coupling is weak in the polaronic frame [which is due to
 small parameter $J_{\perp}/(g\omega) \ll1$], we note that $\rho_T(t) \approx \rho_s(t)\otimes B_0$; consequently,
we obtain
\begin{eqnarray}
 &&\langle 10 | \rho_s^O(t)|01\rangle
\nonumber \\
&&\approx \langle 10 | \sum_n {_{ph}}\langle n|e^{-S}\rho_s(t)\otimes B_0 e^S|n \rangle_{ph}|01\rangle
\nonumber \\
&&\approx \langle 10 | \sum_n {_{ph}}\langle n|e^{-S}\rho_s(t)\otimes 
\sum_m |m\rangle_{ph}~{_{ph}}\langle m| \frac{e^{-\beta\omega_m}}{Z}e^S|n \rangle_{ph}|01\rangle 
\nonumber \\
&&\approx \langle 10 | \rho_s(t) |01\rangle \otimes \nonumber \\
&&~~~~\sum_n {_{ph}}\langle n|e^{-g(a_1-a^{\dagger}_1)}\sum_m |m\rangle_{ph}~{_{ph}}\langle m| \frac{e^{-\beta\omega_m}}{Z}e^{g(a_2-a^{\dagger}_2)}|n \rangle_{ph}
\nonumber \\
&&\approx \langle 10 | \rho_s(t) |01\rangle \otimes 
\sum_m {_{ph}}\langle m| \frac{e^{-\beta\omega_m}}{Z}e^{g(a_2-a^{\dagger}_2)}e^{-g(a_1-a^{\dagger}_1)} |m\rangle_{ph} .
\nonumber \\
\end{eqnarray}
Thus, at 0 {\rm K}, the reduced density matrices in the two frames are related as
$\langle 10 | \rho_s^O(t)|01\rangle = e^{-g^2}\langle 10 | \rho_s(t)|01\rangle$.
}}

{{Before closing this section, we would like to mention that it would be appropriate to treat
each dot as a multi-mode phonon bath that is coupled to a spin; details of such a treatment are given in Appendix C}}. 
 
\section{Discussion and Conclusion}
A system of two spins will have global (local) coupling when the correlation length of the environment is
much larger (smaller) than the distance between the spins; whereas, for correlation length comparable to inter-spin distance, 
local as well as global couplings are relevant.

{In this paper, for a local phonon environment, we have shown that 
a two-spin Heisenberg system 
can coherently retain quantum information encoded in a superposition of dressed spin (polaron) basis
states {{when initially the system and the environment form a product state in the polaronic frame}.
The two-spin system [described by equation (\ref{heisenberg})] undergoes decoherence} 
when subjected to non-Markovian dynamics 
whereas no decoherence results under Markovian evolution. The Heisenberg spin system with $S_T^z=0$, leads to the coupled spins
flipping  simultaneously; consequently, as can be seen from Eq. (\ref{dom_2}), the dominant second-order perturbation process will correspond to
 simultaneous flipping of adjacent spins followed by flipping back to original spin configuration  
with the energy scale for the entire process being
$\frac{J^2_\perp}{g^2 \omega}$ \cite{Pankaj,sahinur}. 
Therefore, to fulfill the condition $\tau_s\gg \tau_c$ for a Markov process, one {\em actually} requires
that  $\frac{J^2_\perp}{g^2 \omega}\ll \omega$. Thus, this condition implies that 
 the long time values given by equations (\ref{expo}) and (\ref{expo1}) show negligible decoherence
in the strong coupling regime ($g^2 \gg 1$).
{It is of interest to note that, consistent with our results, a study of non-equilibrium dynamics of
Holstein polarons (driven by a weak external electric field) in one-dimension 
 reveals that weakly-damped
nearly-adiabatic evolution occurs within
the polaron band  at strong coupling \cite{trugman}.}

In the global phonon case, though the system eigenstates with same $S^z_T$ values form a DFS, decoherence occurs when one considers states 
with different $S^z_T$ values. 

To exercise quantum control, we propose a strategy to suppress decoherence through simultaneously flipping both the spins with externally 
applied fast train of pulses. 
{{One can notice that, for $N$ steps of the evolution described by equation (\ref{U2}),
the system is free of decoherence with the error (up to 
leading order in the small period $\delta t =t/N$) being at most $\delta t ^4$ [see Eq. (\ref{dt3})];
thus the error can be made arbitrarily small by choosing sufficiently large values of $N$.}}

Here, we should mention that the decoherence analysis holds  for optical phonons which naturally appear in many materials
such as the transition metal oxides. For acoustic phonons, the condition 
$J_\perp e^{-g^2} \ll \omega$ can also be satisfied for small systems as the smallest permitted
wave vector is inversely proportional to the system size.

{{
Lastly, we proposed a charge qubit (to be realized in an oxide double quantum dot) that exploits the rich oxide physics
and can possibly meet future challenges of decoherence-free and dissipationless operations
as well as miniaturization.}}

\section{Acknowledgments}
One of the authors (S. Y.) would like to thank 
P. B. Littlewood,  
 T. V. Ramakrishnan,  Sibasish Ghosh, and Atindra Nath Das for valuable discussions.

\section{Appendix A}
We will now obtain the long-time ($t \rightarrow \infty$) behavior of the matrix elements
of $\rho_s(t)$. Using equation (\ref{trick}) in the main text and considering that the initial matrix elements are all real, we have 
[from equations (\ref{sol_offdiag1}), (\ref{sol_offdiag2}), (\ref{sol_diag1}),  and (\ref{sol_diag2})]
\begin{eqnarray}
 &&|\langle \varepsilon_s | \rho_s(t)|\varepsilon_t\rangle|= 
 \langle \varepsilon_s | \rho_s(0)|\varepsilon_t\rangle ~\exp \Bigg[-2K \sum_{m_1,m_2}^{\prime} \frac{C_{m}} 
 {\omega_{m}^2}\Bigg],  
\nonumber \\
 \label{factor}
\end{eqnarray}
\begin{eqnarray} 
 &&\langle \varepsilon_s | \rho_s(t)|\varepsilon_s\rangle  \nonumber \\
 &&= 
 \frac{1}{2}  \langle \varepsilon_s | \rho_s(0)|\varepsilon_s\rangle
  \Big\{1+\exp\Big[-2K \sum_{|m_1 - m_2|= {\rm odd}} \frac{C_{m}}{\omega_{m}^2} \Big]\Big\} \nonumber \\
 &&+\frac{1}{2}  \langle \varepsilon_t | \rho_s(0)|\varepsilon_t\rangle
  \Big\{1-\exp\Big[-2K \sum_{|m_1 - m_2|= {\rm odd}} \frac{C_{m}}{\omega_{m}^2} \Big]\Big\} ,
\nonumber \\
\end{eqnarray}
 and
\begin{eqnarray} 
   &&\langle \varepsilon_t | \rho_s(t)|\varepsilon_t\rangle  \nonumber \\
 &&=\frac{1}{2}  \langle \varepsilon_s | \rho_s(0)|\varepsilon_s\rangle
  \Big\{1-\exp\Big[-2K \sum_{|m_1 - m_2|= {\rm odd}} \frac{C_{m}}{\omega_{m}^2} \Big]\Big\} \nonumber \\
 &&+\frac{1}{2}  \langle \varepsilon_t | \rho_s(0)|\varepsilon_t\rangle
  \Big\{1+\exp\Big[-2K \sum_{|m_1 - m_2|= {\rm odd}} \frac{C_{m}}{\omega_{m}^2} \Big]\Big\} .\nonumber \\
  \end{eqnarray}
In the above equations, for large values of $g$, we will now simplify
the exponential terms  $\exp \Bigg[-2K \sum_{m_1,m_2}^{\prime} \frac{C_{m}} 
 {\omega_{m}^2}\Bigg]$ and $\exp\Big[-2K \sum_{|m_1 - m_2|= {\rm odd}} \frac{C_{m}}{\omega_{m}^2} \Big]$. Let 
 \begin{equation}
  F= \sum_{m_1,m_2}^{\prime}\frac{C_{m}} 
 {\omega_{m}^2}= \frac{1}{\omega^2}\sum_{m_1, m_2}^{\prime} \frac{g^{2(m_1 + m_2)}}{m_1 ! m_2 ! (m_1 + m_2)^2}.
 \end{equation}
On taking double derivative of $F$, one can write
\begin{equation}
{\omega^2} g^2\frac{\partial}{\partial{g^2}}\Big( g^2\frac{\partial{F}}{\partial{g^2}}\Big)=\sum_{m_1, m_2}^{\prime} 
 \frac{g^{2(m_1 + m_2)}}{m_1 ! m_2 ! }=e^{2g^2}-1.
\end{equation}
Next, by integrating the above equation twice (for large values of $g$), we can write
\begin{equation}
{\omega^2} F\approx\frac{e^{2g^2}}{4g^4} ,
\end{equation}
and thus
\begin{equation}
\label{expo}
 \exp \Bigg[-2K \sum_{m_1,m_2}^{\prime} \frac{C_{m}} 
 {\omega_{m}^2}\Bigg]=\exp \Big[-\frac{1}{4g^2}\big(\frac{J_\perp}{g\omega}\big)^2 \Big].
\end{equation}
Similarly, defining
\begin{eqnarray}
 { F'}&&{=} \sum_{|m_1 - m_2|={\rm odd}}\frac{C_{m}} {\omega_{m}^2} \nonumber \\
 &&= \frac{1}{\omega^2}\sum_{|m_1 - m_2|={\rm odd}} \frac{g^{2(m_1 + m_2)}}{m_1 ! m_2 ! (m_1 + m_2)^2},
\end{eqnarray}
we obtain
\begin{eqnarray}
{\omega^2}
 g^2\frac{\partial}{\partial{g^2}}\Big( g^2\frac{\partial{F'}}{\partial{g^2}}\Big)&=&\sum_{|m_1 - m_2|={\rm odd}}  \frac{g^{2(m_1 + m_2)}}{m_1 ! m_2 ! } \nonumber\\
 &=&\sum_{m_1={\rm even}, m_2={\rm odd}}\frac{g^{2(m_1 + m_2)}}{m_1 ! m_2 ! }\nonumber \\
 &&+\sum_{m_1={\rm odd}, m_2={\rm even}}\frac{g^{2(m_1 + m_2)}}{m_1 ! m_2 ! } \nonumber \\
 &=& \frac{1}{2}(e^{2g^2} - e^{-2g^2}) ,
\end{eqnarray}
which on integration yields
\begin{eqnarray}
 {\omega^2} F'&\approx& \frac{e^{2g^2}}{8g^4}.
 \end{eqnarray}
 Therefore, we can write (at large values of $g$)
 \begin{eqnarray}
 \label{expo1}
  \exp \Bigg[-2K \sum_{|m_1 - m_2|={\rm odd}} \frac{C_{m}} 
 {\omega_{m}^2}\Bigg]=\exp \Big[-\frac{1}{8g^2}\big(\frac{J_\perp}{g\omega}\big)^2 \Big].\nonumber \\
 \end{eqnarray}
Here, the important point is to realize that  the effective system behavior, up to second-order in  perturbation theory,
has the energy scale $\frac{J^2_\perp}{g^2 \omega}$ associated with the dominant process (involving the adjacent spins flipping simultaneously
and then reverting back to the original spin state simultaneously)
 \cite{Pankaj,sahinur}. So, the condition $\gamma^2 = (\frac{J^2_\perp}{g^2 \omega}) / \omega \ll 1$ leads to the 
condition $\tau_s \gg \tau_c$ for Markovian dynamics.
It is evident from the equations (\ref{longtime}), (\ref{longtime1}), and (\ref{longtime3})  that, with decreasing 
$\gamma$, the decoherence becomes less; this  means {that the evolutionary process} becomes closer to the Markovian dynamics.
\section{\bf{Appendix B: Non-Markovian dynamics including the phase factor due to energy gap of system}}
{{Here, we consider the phase factor due to the energy gap of the system. We show below that, although there are additional oscillatory features 
in the coherence  factor and in the population difference due to the presence of this phase factor,
 the amount of decoherence does not change much. For the population difference [$P_d (t)$], there are some interesting situations such as,
 if one starts with zero value of the initial population difference, $P_d(t)$ changes by a small amount 
 at later times [whereas there is no change if the system excitation is neglected (see Sec. IV)].}}
\subsection{\bf Off-diagonal elements}
We will first analyze the off-diagonal elements for the effect of the phase term due to the energy difference $\varepsilon_s -\varepsilon_t$.  
{
Starting with equation (\ref{mas}), at {\rm T = 0 K}, one gets the following form for the master equation
\begin{widetext}
\begin{eqnarray}
\frac{d \tilde{\rho}_s(t)}{dt} &=& - \sum_{n} \int_0^t d\tau \left[~ _{ph}\langle 0 |
 \tilde{H}_I^L(t)| n \rangle_{ph}~ _{ph}\langle n | \tilde{H}_I^L(\tau)| 0 \rangle_{ph} \tilde{\rho}_s(t) \right. 
\nonumber \\ 
 &&~~~~~~~~~~~~~~~~~~~ - {~_{ph}}\langle n| \tilde{H}_I^L(t)|0\rangle_{ph}\tilde{\rho}_s(t) {_{ph}}\langle 0 | \tilde{H}_I^L(\tau) |n\rangle_{ph} \nonumber \\
 && ~~~~~~~~~~~~~~~~~~~- {~_{ph}}\langle n| \tilde{H}_I^L(\tau)| 0\rangle_{ph} \tilde{\rho}_s(t) {_{ph}}\langle 0| \tilde{H}_I^L(t) |n\rangle_{ph} 
 \nonumber \\
&& ~~~~~~~~~~~~~~~~~~~
 + \left .
~ \tilde{\rho}_s(t)  _{ph}\langle 0 |
 \tilde{H}_I^L(\tau)| n \rangle_{ph}~ _{ph}\langle n | \tilde{H}_I^L(t)| 0 \rangle_{ph} 
 \right] .\nonumber \\
\label{mas0}
\end{eqnarray}
On taking the matrix elements (with respect to $|\varepsilon_s\rangle$ and $|\varepsilon_t\rangle$) on both sides of equation (\ref{mas0}), we
get the following coupled differential equations:
\begin{eqnarray}
 \frac{d \langle \varepsilon_s|\tilde{\rho}_s(t)|\varepsilon_t \rangle}{dt} = -K\Bigg[&&\langle \varepsilon_s|\tilde{\rho}_s(t)|\varepsilon_t \rangle
 \Bigg(2 \sum^{\prime}_{|n_1 - n_2|={\rm even},0}C_{n} \frac{\sin(\omega_{n}t)}{\omega_{n}} \nonumber \\
 &&-\frac{i}{2}\sum^{\prime}_{|n_1 - n_2|={\rm odd}}
C_{n}\Big \{\frac{2\Delta \varepsilon}{\omega^2_{n}-\Delta \varepsilon^2}+e^{i\Delta \varepsilon t}\Big(\frac{e^{i\omega_{n}t}}{\omega_{n}
+\Delta \varepsilon}-\frac{e^{-i\omega_{n}t}}{\omega_{n}
-\Delta \varepsilon}\Big)\Big\}\Bigg) \nonumber \\
&&+\langle \varepsilon_t|\tilde{\rho}_s(t)|\varepsilon_s \rangle \frac{i}{2}\sum^{\prime}_{|n_1 - n_2|={\rm odd}}  {C_{n}}
\Big \{e^{2i\Delta \varepsilon t}
\frac{2\Delta \varepsilon}{\omega^2_{n}-\Delta \varepsilon^2}+e^{i\Delta \varepsilon t}\Big(\frac{e^{-i\omega_{n}t}}{\omega_{n}
+\Delta \varepsilon}-\frac{e^{i\omega_{n}t}}{\omega_{n}
-\Delta \varepsilon}\Big)\Big\}\Bigg]  ,
\end{eqnarray}
\end{widetext}
and its complex conjugate equation for $d\langle \varepsilon_t|\tilde{\rho}_s(t)|\varepsilon_s\rangle/dt$, {where $\omega_n = 
 \omega (n_1+n_2)$ and $C_{n}=C_{n_1,n_2}$}. Here $\Delta \varepsilon\equiv 
\varepsilon_s -\varepsilon_t=J_{\perp}e^{-g^2}$ 
is the system excitation energy. We solve these coupled differential equations numerically and plot 
the coherence factor {${\rm C(t)}$} in  Fig. (\ref{fig3}). 
Compared to the case (depicted in Fig. (1)) in the main text, there is an additional oscillatory feature due to the presence of the  excitation
energy $\Delta \varepsilon$; nevertheless, the decoherence is 
still quite small.
\begin{figure}
\centerline{\includegraphics[angle=90,angle=90,angle=90,width=3.5in,height=3in]{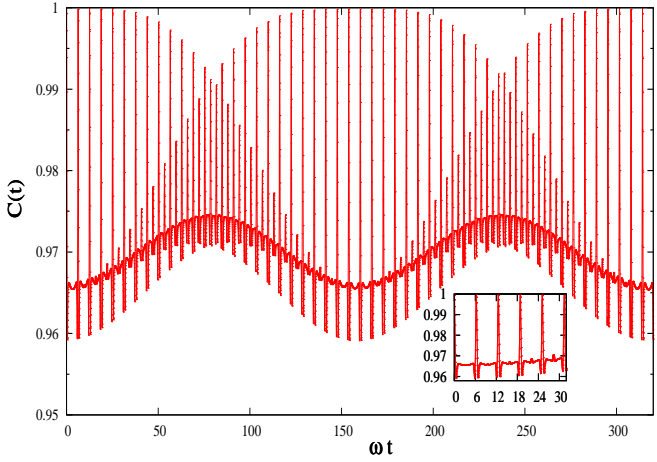}}
\caption{Time variation of coherence factor for $g=2$ and $\frac{\Delta \varepsilon}{\omega}=0.02$. The inset shows
that {${\rm C(t)}$} is similar to Fig. (1) for $\omega t/ 2 \pi \sim 1$.}
\label{fig3}
\end{figure}
In Fig. (\ref{fig3}), since $\omega=50\Delta \varepsilon$, one full period corresponds to the time period $t$ satisfying
 the condition $\Delta \varepsilon t=2 \pi$, i.e., $\omega t = 2 \pi \omega/\Delta \varepsilon = 50 \times 2 \pi$.
 To explain the oscillatory behavior of Fig. (\ref{fig3}), because $\frac{\Delta \varepsilon}{\omega}\ll 1$,
one  can safely neglect $\Delta \varepsilon$ compared to $\omega$ 
 everywhere except in the phase factors and get the differential equations for the real and the imaginary part 
of $\langle \varepsilon_s|\tilde{\rho}_s(t)|\varepsilon_t\rangle$.
\begin{widetext}
\begin{eqnarray}
\frac{d \big(\langle \varepsilon_s|\tilde{\rho}_s(t)|\varepsilon_t \rangle+\langle \varepsilon_t|\tilde{\rho}_s(t)|\varepsilon_s\rangle\big)}{dt}
=-2K\big(\langle \varepsilon_s|\tilde{\rho}_s(t)|\varepsilon_t \rangle+\langle \varepsilon_t|\tilde{\rho}_s(t)|\varepsilon_s\rangle\big)
\Bigg[&&\sum^{\prime}_{|n_1 - n_2|={\rm even},0}C_{n} \frac{\sin(\omega_{n}t)}{\omega_{n}} \nonumber \\ 
&&+\cos\Delta \varepsilon t 
\sum^{\prime}_{|n_1- n_2|={\rm odd}}C_{n} \frac{\sin(\omega_{n}t)}{\omega_{n}}\Bigg] ,
\label{real}
\end{eqnarray}
and
\begin{eqnarray}
 \frac{d \big(\langle \varepsilon_s|\tilde{\rho}_s(t)|\varepsilon_t \rangle-\langle \varepsilon_t|\tilde{\rho}_s(t)|\varepsilon_s\rangle\big)}{dt}
 =-{{2} } K\Bigg[&&\big(\langle \varepsilon_s|\tilde{\rho}_s(t)|\varepsilon_t \rangle-\langle \varepsilon_t|\tilde{\rho}_s(t)|\varepsilon_s\rangle\big)
 \sum^{\prime}_{|n_1 - n_2|={\rm even},0}C_{n} \frac{\sin(\omega_{n}t)}{\omega_{n}} \nonumber \\
&&+ i\big(\langle \varepsilon_s|\tilde{\rho}_s(t)|\varepsilon_t \rangle+\langle \varepsilon_t|\tilde{\rho}_s(t)|\varepsilon_s\rangle\big)
\sin\Delta \varepsilon t 
\sum^{\prime}_{|n_1- n_2|={\rm odd}}C_{n} \frac{\sin(\omega_{n}t)}{\omega_{n}}\Bigg] .
\label{imaginary}
\end{eqnarray}
From equations (\ref{real}) and (\ref{imaginary}), for real values of $\langle \varepsilon_s|\tilde{\rho}_s(0)|\varepsilon_t \rangle$,
 one can show that the imaginary part of $\langle \varepsilon_s|\tilde{\rho}_s(t)|\varepsilon_t \rangle$ is much smaller than its real part 
for the chosen values of the parameters (i.e., $\Delta \varepsilon/\omega \le 0.02$ and $g > 1$);  thus
 the off-diagonal density matrix element can be well approximated by its real value only. Now, the solution of equation 
(\ref{real}) is given by
\begin{eqnarray}
\!\!\!\! \big(\langle \varepsilon_s|\tilde{\rho}_s(t)|\varepsilon_t \rangle+\langle \varepsilon_t|\tilde{\rho}_s(t)|\varepsilon_s\rangle\big)
 =\big(\langle \varepsilon_s|\rho_s(0)|\varepsilon_t \rangle+\langle \varepsilon_t|\rho_s(0)|\varepsilon_s\rangle\big)
 \exp\Bigg[&&-2K \Big\{\sum^{\prime}_{n_1,n_2} C_{n} \frac{1-\cos(\omega_{n}t)}{\omega^2_{n}} \nonumber \\
 &&+\sum^{\prime}_{|n_1 - n_2|={\rm odd}}C_{n} (1-\cos \Delta \varepsilon t)\frac{\cos(\omega_{n}t)}{\omega^2_{n}}\Big\} \Bigg] .
 \label{realsol}
\end{eqnarray}
\end{widetext}
The exponential term in  equation (\ref{realsol}) can be treated as the coherence factor {C(t)} and the second term in the exponent is the correction due 
to the excitation energy $\Delta \varepsilon$ of the system. Even with the correction term, {C(t)} still peaks at $\omega t=2m\pi$; however
the peak values monotonically decrease from 
$\omega t=0$ to $\omega t=50 \pi$ because the factor $(1-\cos \Delta \varepsilon t)$ in the exponent increases from 0 to 2. 
As $\cos(\omega_{n}t)=-1$ for odd values of $(n_1+n_2)$, at $\omega t=(2m+1)\pi$, the correction term produces peaks in {C(t)}. Moreover, 
because of the factor $(1-\cos \Delta \varepsilon t)$, these peaks increase monotonically from $\omega t=0$ to $\omega t=50 \pi$
 as can be seen in figure (\ref{fig3}). The features 
of the figure (\ref{fig3}), from $\omega t=50 \pi$ to $\omega t=100 \pi$ can be explained by using a similar logic.

All in all, based on Fig. (\ref{fig3}), we can conclude that decoherence does not increase when we include
the phase factor due to the energy gap of the system.
\begin{figure}
\centerline{\includegraphics[angle=90,angle=90,angle=90,width=3.5in,height=3.5in]{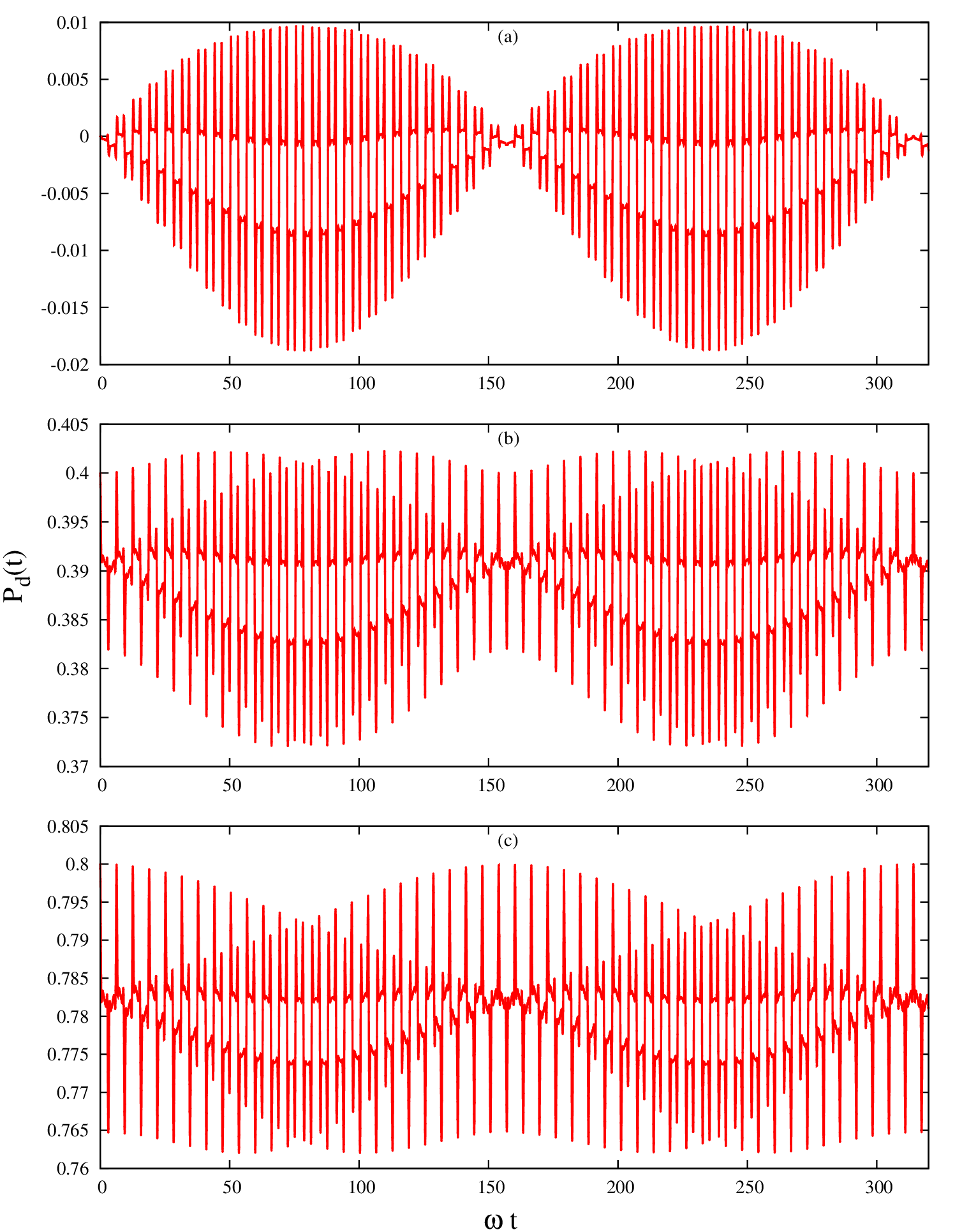}}
\caption{Time variation of the population difference $P_d(t)$ for $g=2$ and $\frac{\Delta \varepsilon}{\omega}=0.02$. Figures (a), (b), and (c) correspond to 
the initial population difference [$P_d(0)$] values 0.0, 0.4, and 0.8, respectively.}
\label{fig4}
\end{figure}
}

\subsection{\bf Diagonal elements}
{{
Similar to the above considerations for the off-diagonal terms, one can obtain the diagonal terms of the density matrix.
 From  equation (\ref{mas0}), one gets the following differential equation for the population difference 
$P_d(t)\equiv \langle \varepsilon_s|{\rho}_s(t)|\varepsilon_s \rangle-\langle \varepsilon_t|{\rho}_s(t)|\varepsilon_t\rangle$:
\begin{widetext}
\begin{eqnarray}
\frac{d P_d(t)}{dt}
=-{{K}}\Bigg[ P_d(t)
\sum^{\prime}_{|n_1- n_2|={\rm odd}} C_{n} \Bigg(\frac{\sin(\omega_{n}+\Delta \varepsilon)t}{\omega_{n}+\Delta \varepsilon}
+\frac{\sin(\omega_{n}-\Delta \varepsilon)t}{\omega_{n}-\Delta \varepsilon}\Bigg )\nonumber \\
~~~~~~~~~~~~~~~-\sum^{\prime}_{|n_1- n_2|={\rm odd}} C_{n}  \Bigg(\frac{\sin(\omega_{n}+\Delta \varepsilon)t}{\omega_{n}+\Delta \varepsilon}
-\frac{\sin(\omega_{n}-\Delta \varepsilon)t}{\omega_{n}-\Delta \varepsilon}\Bigg)\Bigg] .\nonumber \\
\label{pd1}
\end{eqnarray}
We solve the above equation numerically and show the dynamics of $P_d(t)$ in figure (\ref{fig4}). Now, to explain Fig. (\ref{fig4}) 
(as $\frac{\Delta \varepsilon}{\omega}\ll 1$) we neglect $\Delta \varepsilon$ compared to $\omega$  everywhere except in the phase factors 
of equation (\ref{pd1}) and obtain  
\begin{eqnarray}
\frac{d P_d(t)}{dt}
=-2K\Bigg[P_d(t) 
\sum^{\prime}_{|n_1- n_2|={\rm odd}} C_{n} \frac{\sin(\omega_{n} t) \cos\Delta \varepsilon t}{\omega_{n}} \nonumber \\
~~~~~~~~~~~~~~~-\sum^{\prime}_{|n_1- n_2|={\rm odd}} C_{n} \frac{\cos(\omega_{n} t) \sin\Delta \varepsilon t}{\omega_{n}}\Bigg].
\label{pd}
\end{eqnarray}
The solution of  equation (\ref{pd}) can be expressed as
\begin{eqnarray}
 P_d(t)
 =P_d(0)
 e^{-\int_0^t A(t^{\prime}) dt^{\prime}} +  e^{-\int_0^t A(t^{\prime}) dt^{\prime}} \int_0^t  e^{\int_0^{t^{\prime}} A(t^{\prime\prime}) 
 dt^{\prime\prime}} B(t^{\prime}) dt^{\prime}, \nonumber \\
\label{Pd_sol}
 \end{eqnarray}
where $A(t)=2K\sum^{\prime}_{|n_1- n_2|={\rm odd}} C_{n} \frac{\sin(\omega_{n} t) \cos\Delta \varepsilon t}{\omega_{n}}$
and $B(t)=2K\sum^{\prime}_{|n_1- n_2|={\rm odd}} C_{n} \frac{\cos(\omega_{n} t) \sin\Delta \varepsilon t}{\omega_{n}}$.
\end{widetext}
One can see that the above solution contains a homogeneous part dependent on the initial population difference and an inhomogeneous part independent of 
the initial condition. 
When $\Delta \varepsilon=0$, the solution reduces to the case analyzed in  Sec. IV with $P_d(t)$ being given by
\begin{eqnarray}
\!\!\!\!\!\!\!\!\! P_d(t)
 &=&P_d(0)
 e^{-2K\big [\sum^{\prime}_{|n_1- n_2|={\rm odd}}{{C_{n}}} \frac{1-\cos(\omega_{n} t)}{\omega^2_{n}}\big]}.
\end{eqnarray}
In Eq. (\ref{Pd_sol}), for zero value of $P_d(0)$,  only the inhomogeneous part [shown in  figure (\ref{fig4}a)] will contribute
{{and $P_d(t)$ changes by a small amount 
 at later times.}} The inhomogeneous term 
vanishes if $\Delta \varepsilon=0$; thus, for $\Delta \varepsilon=0$ and initial value $P_d(0)=0$, the value of $P_d(t)$  
(identical to the case in Sec. IV) will remain zero for all times $t$.
The factor $H(t) \equiv e^{-\int_0^t A(t^{\prime}) dt^{\prime}}$, in the homogeneous term of Eq. (\ref{Pd_sol}), can be written as 
\begin{eqnarray}
\!\!\!\!\!\! e^{-\int_0^t A(t^{\prime}) dt^{\prime}}
= e^{-2K\big [ \sum^{\prime}_{|n_1- n_2|={\rm odd}}
{{C_{n}}} \frac{1-\cos(\omega_{n} t)\cos \Delta \varepsilon t}{\omega^2_{n}}\big ]} ,
\nonumber \\
 \label{hom}
\end{eqnarray}
and is plotted in figure (\ref{fig5}).
\begin{figure}
\centerline{\includegraphics[angle=90,angle=90,angle=90,width=3.5in,height=2.5in]{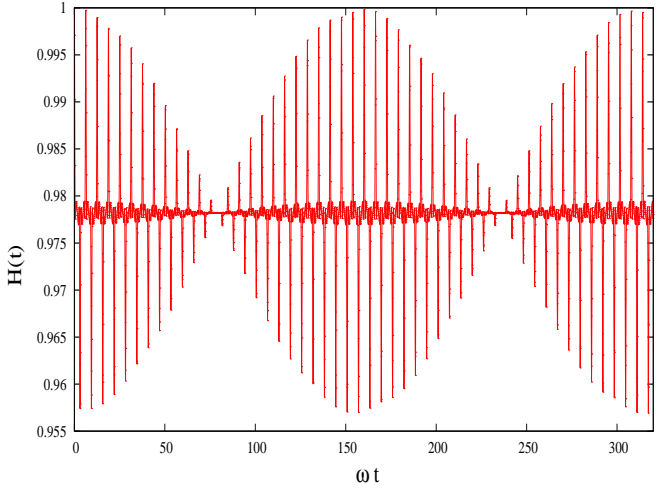}}
\caption{Time variation of the factor $H(t)= e^{-\int_0^t A(t^{\prime}) dt^{\prime}}$ for $g=2$ and $\frac{\Delta \varepsilon}{\omega}=0.02$. }
\label{fig5}
\end{figure}
One can explain the features of figure (\ref{fig5}) by analyzing the exponent in equation (\ref{hom}). 
The $\cos(\omega_{n} t)$ term in the exponent attains the maximum value of 1 at $\omega t=2m\pi$ and the minimum value of -1 at $\omega t=(2m+1)\pi$.
Furthermore, as $\Delta \varepsilon t$ increases from $0$ to $\pi$ (i.e., as $\omega t$ increases from $0$ to $50 \pi$), the value of 
$\cos(\Delta \varepsilon t)$ decreases from 1 to -1. Thus, as $\omega t$ increases from $0$ to $50 \pi$, there is
a monotonic decrease (increase) in the value of the local
extrema at $\omega t=2m\pi$ [$\omega t=(2m+1)\pi$] for the term
$ \cos(\omega_{n} t)\cos \Delta \varepsilon t$ in the exponent; consequently, the  same trend is reflected for $H(t)$ 
in the region $0 \le \omega t \le 50 \pi$.   
 The remaining part of the figure (\ref{fig5}) (i.e., for $50 \pi \le \omega t \le 100 \pi$) can be explained  using a similar logic.}}

{{The inhomogeneous term in Eq. (\ref{Pd_sol}) can be simplified by performing integration by parts and can be written as 
\begin{widetext}
\begin{eqnarray}
&& e^{-\int_0^t A(t^{\prime}) dt^{\prime}} \int_0^t  e^{\int_0^{t^{\prime}} A(t^{\prime\prime})dt^{\prime\prime}} 
 B(t^{\prime}) dt^{\prime} \nonumber \\
 &=& {{2K  \Bigg[\sin\Delta \varepsilon t \sum^{\prime}_{|n_1-n_2|={\rm odd}} C_{n} \frac{\sin(\omega_{n} t)}{\omega^2_{n}}}}
  \nonumber \\
&&~~~~~~~~~~~~~~~~~~~~~
{{-e^{-\int_0^t A(t^{\prime})dt^{\prime}}
\int_0^t [\sin\Delta \varepsilon t^{\prime}A(t^{\prime})+\Delta \varepsilon \cos\Delta \varepsilon t^{\prime}]e^{\int_0^{t^{\prime}} A(t^{\prime\prime}) dt^{\prime\prime}}
\sum^{\prime}_{|n_1-n_2|={\rm odd}} C_{n} \frac{\sin(\omega_{n} t^{\prime})}{\omega^2_{n}}
  dt^{\prime}
\Bigg]}} . \nonumber \\
\label{partial}
 \end{eqnarray}
\end{widetext}
 Now, figure (\ref{fig6}) shows the plot of the first term [$I(t)$] on the right-hand side of equation (\ref{partial});
the figure indicates  splitting up of $I(t)$ into positive and negative regions for values of $\omega t$ in
the regions $2m\pi < \omega t < (2m+1)\pi $ and
$(2m+1) \pi < \omega t < (2m+2) \pi$, respectively.
A similar splitting of the total inhomogeneous term [i.e., sum of all the terms of equation (\ref{partial})]
is reflected in Fig. (\ref{fig4} a); however, there is an asymmetry in the splitting  which results when we consider all the
 terms on the right-hand
side of Eq. (\ref{partial}).
 At $\omega t=m\pi$ all the $\sin(\omega_{n}t)$ terms are zero [see Fig. (\ref{fig6})]; for  $2m\pi < \omega t < (2m+1)\pi $ 
[$(2m+1) \pi < \omega t < (2m+2) \pi$] the  $\sin(\omega_{n}t)$ terms are positive (negative) and interfere destructively
to produce a flat region away from the edges (i.e., $\omega t = n\pi t$). 
Additionally, as $\Delta \varepsilon t$ increases from 0 to $\frac{\pi}{2}$ [i.e.,  $\omega t$ increases from $0$ to $25\pi$],
the prefactor $\sin \Delta \varepsilon t$ in $I(t)$
increases from 
 0 to 1
 and 
after that as  $\Delta \varepsilon t$ increases to $\pi$  (i.e., $\omega t$ increases to $50\pi$)
the $\sin \Delta \varepsilon t$ term decreases to zero. Consequently, the splitting in $I(t)$
is maximum at $\omega t=25\pi, 75\pi$ and minimum at $\omega t=0, 50\pi, 100\pi$.
 \begin{figure}
\centerline{\includegraphics[angle=90,angle=90,angle=90,width=3.5in,height=2.5in]{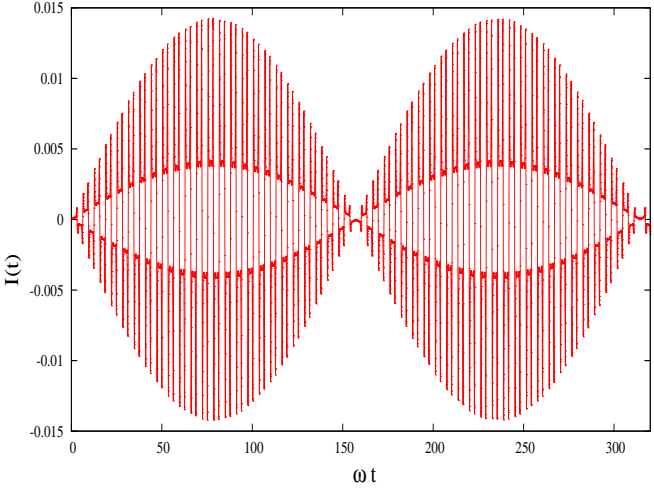}}
\caption{Time variation of the first term in the inhomogeneous factor [expressed in Eq. (\ref{partial})]
{{for $g=2$ and $\frac{\Delta \varepsilon}{\omega}=0.02$}}. }
\label{fig6}
\end{figure}
{
\section {Appendix C: Spins interacting with 
 multi-mode local phonons}}
Here we consider the case when spins interact with  momentum ($k$) dependent local phonon modes.
The phonon bath Hamiltonian is given by $H_{B} =  \sum_{i,k}\omega_k a_{i,k}^{\dagger} a_{i,k}$ while
the system-bath interaction Hamiltonian is of the form
$ \sum_{i,k}g_k \omega_k S_{i}^{z}( a_{i,k} + a_{i,k}^{\dagger})$. Upon performing the Lang-Firsov
transformation $H^L = e^S H e^{-S}$ where $S= -\sum_{i,k}g_k S^z_i (a_{i,k} - a_{i,k}^{\dagger})/\sqrt{N}$,
in the Lang-Firsov frame,
we get the system Hamiltonian to be
\begin{equation}
 H_s^{L}= J_\lVert S_{1}^{z}S_{2}^{z} - \frac{J_\perp e^{-\frac{1}{N}\sum_k g_k^2}}{2} (S_{1}^{+}S_{2}^{-}+S_{2}^{+}S_{1}^{-}),
\end{equation}
 the bath term to be
\begin{eqnarray}
 H_{B}^L = \sum_{i,k} \omega_k a_{i,k}^{\dagger} a_{i,k} ,
\end{eqnarray}
and
the interaction Hamiltonian to be
\begin{eqnarray}
 H_I^{L}&=& -\frac{1}{2} [J_\perp ^+ S_{1}^{+}S_{2}^{-} + J_\perp ^- S_{2}^{+}S_{1}^{-} ],
\end{eqnarray}
where 
\begin{eqnarray}
 J_\perp ^{\pm} &=& J_\perp e^{\pm \frac{1}{\sqrt{N}}\sum_{k}~ g_k [(a_{2,k} - a_{2,k}^{\dagger})-(a_{1,k} - a_{1,k}^{\dagger})]} - 
 J_\perp e^{-\frac{1}{{N}}\sum_{k}g_k^2} .
 \nonumber \\
\end{eqnarray}
Now, we wish to study the dynamical evolution of the density matrix by employing the {\rm T = 0 K} master equation 
given by Eqn. (\ref{mas0}).
To this end, we calculate the matrix element $_{ph}\langle \{0_1^k\},\{0_2^k\}|H_I^{LF}|\{m_1^k\},\{m_2^k\}\rangle_{ph}$ 
[with $m_i^k$ ($0_i^k$) being the number (vacuum) of phonons in mode $k$ at site $i$] and get the expression
\begin{eqnarray}
\!\! &&_{ph}\langle \{0_1^k\},\{0_2^k\}|H_I^{LF}|\{m_1^k\},\{m_2^k\}\rangle_{ph} \nonumber \\
&=&\frac{J_\perp}{2} e^{-\frac{1}{{N}}\sum_{k}g_k^2} \Bigg(\prod_k \frac{(\frac{g_k}{\sqrt{N}})^{(m_1^k+m_2^k)}}{\sqrt{m_1^k!m_2^k!}}
 \Bigg)
 \nonumber \\
 && \times(-1)^{\sum_{k} m_1^k} \big[S_{1}^{+}S_{2}^{-}+(-1)^{\sum_{k} (m_1^k-m_2^k)}S_{2}^{+}S_{1}^{-}\big].
\nonumber \\
\end{eqnarray}
Using the above result in the master equation [of Eq. (\ref{mas0})] and taking matrix element with respect to $|\varepsilon_s\rangle$ and 
$|\varepsilon_t\rangle$, we get 
\begin{eqnarray}
  \nonumber &&\frac{d \langle \varepsilon_s | \tilde{\rho}_s(t)|\varepsilon_t\rangle }{dt} \nonumber \\
  &&= -\bar{K} \Bigg[ \sum^\prime_{\{m^k\}}~\bar{C}_{m} 
  \langle \varepsilon_s | \tilde{\rho}_s(t)|\varepsilon_t\rangle \frac{\sin(\bar{\omega}_{{m}}t)}{\bar{\omega}_{{m}}} \nonumber  \\
 &&~~~~~~  + \sum_{|\sum_{k}(m^k_1- m^k_2)| = {\rm odd}} \bar{C}_{{m}} 
  \langle \varepsilon_t | \tilde{\rho}_s(t)|\varepsilon_s\rangle \frac{\sin(\bar{\omega}_{{m}}t)}{\bar{\omega}_{{m}}} \nonumber \\
  &&~~~~~~+ \sum^{\prime}_{|\sum_{k}(m^k_1- m^k_2)| = {\rm even},0} \bar{C}_{{m}} 
  \langle \varepsilon_s | \tilde{\rho}_s(t)|\varepsilon_t\rangle \frac{\sin(\bar{\omega}_{{m}}t)}{\bar{\omega}_{{m}}}\Bigg],\nonumber \\
  \label{multimas}
 \end{eqnarray}
 where $\sum^\prime_{\{m^k\}}$ \big ($\sum^{\prime}_{|\sum_{k}(m^k_1- m^k_2)| = {\rm even},0}$ \big) implies summation over
{{elements of  $\{m_k\} \equiv \{m_1^k\} \cup \{m_2^k\} $}}   
\big (0 or even values of $|\sum_{k}(m^k_1- m^k_2)|$ \big ) excluding the case when $\{m_1^k\}=\{0_1^k\}$ and $\{m_2^k\}=\{0_2^k\}$.
 Furthermore, $\bar{\omega}_{{m}}\equiv\sum_{k} \omega_k (m_1^k+m_2^k)$ is the 
 eigen energy of the phonon state $|\{m_1^k\},\{m_2^k\}\rangle_{ph}$, $\bar{K}\equiv\frac{J^2_\perp}{2} e^{-\frac{2}{N}\sum_{k}g_k^2}$, and 
 $\bar{C}_m\equiv\prod_k \frac{(\frac{g_k}{\sqrt{N}})^{2(m_1^k+m_2^k)}}{{m_1^k!m_2^k!}}$. In obtaining
the above Eq. (\ref{multimas}), we ignored $|\varepsilon_s -\varepsilon_t|$ compared to the much larger $\bar{\omega}_{{m}}$.
The equation (\ref{multimas}) 
has a form similar to that of equation (\ref{sz0}) and 
 hence the solutions of equation (\ref{multimas}) and its complex conjugate are given by
 \begin{eqnarray}
  &&\langle \varepsilon_s | \rho_s(t)|\varepsilon_t\rangle e^{i(\varepsilon_s-\varepsilon_t)t} \nonumber \\
  &&= 
  \frac{1}{2}\Bigg[ \langle \varepsilon_s | \rho_s(0)|\varepsilon_t\rangle \Big\{\exp[-2\bar{K} \bar{D}_{\rm all}(t)] \nonumber \\ 
  &&~~~~~~~~~~~~~~~~~~~~~~+\exp[-2\bar{K} \bar{D}_{\rm even}(t)] \Big\} \nonumber \\ 
  &&~~~+\langle \varepsilon_t | \rho_s(0)|\varepsilon_s\rangle \Big\{\exp[-2\bar{K} \bar{D}_{\rm all}(t)] \nonumber \\ 
  &&~~~~~~~~~~~~~~~~~~~~~~~~-\exp[-2\bar{K} \bar{D}_{\rm even}(t) ] \Big\} 
\nonumber  \Bigg], \\
\label{1sol_offdiag1}
\end{eqnarray}
and
\begin{eqnarray}
  &&\langle \varepsilon_t | \rho_s(t)|\varepsilon_s\rangle e^{i(\varepsilon_t-\varepsilon_s)t} \nonumber \\
  &&=
  \frac{1}{2}\Bigg[ \langle \varepsilon_t | \rho_s(0)|\varepsilon_s\rangle \Big\{\exp[-2\bar{K} \bar{D}_{\rm all}(t)] \nonumber \\ 
  &&~~~~~~~~~~~~~~~~~~~~~~+\exp[-2\bar{K} \bar{D}_{\rm even}(t)] \Big\} \nonumber \\ 
  &&~~~+\langle \varepsilon_s | \rho_s(0)|\varepsilon_t\rangle \Big\{\exp[-2\bar{K}  \bar{D}_{\rm all}(t)] \nonumber \\ 
  &&~~~~~~~~~~~~~~~~~~~~~~~~-\exp[-2\bar{K}\bar{D}_{\rm even}(t)] \Big\} \nonumber \Bigg], \\
  \label{1sol_offdiag2}
\end{eqnarray}
where we define
\begin{eqnarray}
 \bar{D}_{\rm all}(t)&=&\sum^{\prime}_{\{m^k_1\}, \{m^k_2\}}\bar{C}_{m} 
  \frac{(1-\cos(\bar{\omega}_{m}t))}{\bar{\omega}_{m}^2}, \label{dfull} \\
  \bar{D}_{\rm even}(t)&=&\sum^{\prime}_{|\sum_{k}(m^k_1- m^k_2)| = {\rm even},0} \bar{C}_{m} 
  \frac{(1-\cos(\bar{\omega}_{m}t))}{\bar{\omega}_{m}^2}, \label{deven} \nonumber \\
  &&\!\!\!\!\!\!\!\!\!\!\!\!\!\!\!\!\!\!\!\!\!\!\!\!\!\!\!\!\!\!\!\!\!\!\!\!\!\!\!\!\!\!\!\!\!\!\!\!\!\!\!\!\
{\rm and}  \\
  \bar{D}_{\rm odd}(t)&=&\sum^{\prime}_{|\sum_{k}(m^k_1- m^k_2)| = {\rm odd}} \bar{C}_{m} 
  \frac{(1-\cos(\bar{\omega}_{m}t))}{\bar{\omega}_{m}^2}.
  \label{dodd}
  \end{eqnarray}
 Following a similar procedure, we get the following solution for the 
population  difference: 
 \begin{eqnarray}
  && \!\!\!\!\!\! \big(\langle \varepsilon_s|{\rho}_s(t)|\varepsilon_s\rangle-\langle \varepsilon_t|{\rho}_s(t)|\varepsilon_t\rangle\big) \nonumber \\
 && =\big(\langle \varepsilon_s|{\rho}_s(0)|\varepsilon_s\rangle-\langle \varepsilon_t|{\rho}_s(0)|\varepsilon_t\rangle\big)
  \exp[-2\bar{K}\bar{D}_{\rm odd}(t)]. \nonumber \\
  \label{1sol_diag}
 \end{eqnarray}
For real values of $\langle \varepsilon_s | \rho_s(0)|\varepsilon_t\rangle$, {{the solution for the off-diagonal element}}
can be written in the simpler form
\begin{eqnarray}
\!\!\!\! |\langle \varepsilon_s | \rho_s(t)|\varepsilon_t\rangle|&=& |\langle \varepsilon_s | \rho_s(0)|\varepsilon_t\rangle| \exp[-2\bar{K} \bar{D}_{\rm all}(t)].
\end{eqnarray}
Now, we simplify the exponential factor $\exp[-2\bar{K} \bar{D}_{\rm all}(t)]$ (appearing in the
above equation) as follows:
\begin{widetext}
\begin{eqnarray}
\!\!\!\!\!\!\!\! \exp [-2\bar{K} \bar{D}_{\rm all}(t)]
&=& \exp \Bigg [-\bar{K} \sum^{\prime}_{\{m_1^k\},\{m_2^k\}}\Bigg (\prod_k \frac{(\frac{g_k}{\sqrt{N}})^{2(m_1^k+m_2^k)}}{m_1^k!m_2^k!} \Bigg ) 
 \int_0^t dt^{\prime}\int_0^{t^{\prime}} dt^{\prime\prime}(e^{i\bar{\omega}_mt^{\prime\prime}}+e^{-i\bar{\omega}_m t^{\prime\prime} } )\Bigg ] \nonumber \\
 &=&\exp \Bigg [-\bar{K} \int_0^t dt^{\prime}\int_0^{t^{\prime}} dt^{\prime\prime}\sum^{\prime}_{\{m_1^k\},\{m_2^k\}}\prod_k \Bigg (\frac{(\frac{g_k}{\sqrt{N}})^{2(m_1^k+m_2^k)}}{m_1^k!m_2^k!}
e^{i{\omega}_k(m_1^k+m_2^k)t^{\prime\prime}} \Bigg) \Bigg ] \nonumber \\
&&\times \exp\Bigg [-\bar{K} \int_0^t dt^{\prime}\int_0^{t^{\prime}} dt^{\prime\prime}\sum^{\prime}_{\{m_1^k\},\{m_2^k\}}\prod_k \Bigg (\frac{(\frac{g_k}{\sqrt{N}})^{2(m_1^k+m_2^k)}}{m_1^k!m_2^k!}
e^{-i{\omega}_k(m_1^k+m_2^k)t^{\prime\prime}} \Bigg ) \Bigg]\nonumber \\
&=&\exp\Bigg[-\bar{K} \int_0^t dt^{\prime}\int_0^{t^{\prime}} dt^{\prime\prime}\bigg[e^{\frac{2}{N}\sum_{k}g_k^2 e^{i\omega_k t^{\prime\prime}}}-1 \bigg]\Bigg]
\exp\Bigg[-\bar{K} \int_0^t dt^{\prime}\int_0^{t^{\prime}} dt^{\prime\prime} \bigg[e^{\frac{2}{N}\sum_{k}g_k^2 e^{-i\omega_k t^{\prime\prime}}}-1\bigg]\Bigg]\nonumber \\
&=&\exp\Bigg[-2\bar{K}\int_0^t dt^{\prime}\int_0^{t^{\prime}} dt^{\prime\prime}
\bigg[\exp\bigg(\frac{2}{N\pi}\int_0^\infty \frac{J(\omega)}{\omega^2}\cos\omega t^{\prime\prime}d\omega\bigg)
\cos\bigg (\frac{2}{N\pi}\int_0^\infty \frac{J(\omega)}{\omega^2}\sin\omega t^{\prime\prime}d\omega \bigg )-1\bigg]\Bigg].
\nonumber \\
\label{multi_offdiag}
\end{eqnarray}      
\end{widetext}
Here, the spectral function for the phonon bath $J(\omega)=\pi\sum_{k}g_k^2 \omega_k^2 \delta (\omega-\omega_k)$ accounts for 
the coupling between the system and 
the various phonon modes. 
Next, we simplify the factor $\exp[-2\bar{K} \bar{D}_{\rm odd}(t)]$ associated with the population difference given by Eq. (\ref{1sol_diag}).
\begin{widetext}
\begin{eqnarray}
 &&\exp[-2\bar{K} \bar{D}_{\rm odd}(t)]
\nonumber \\
 &&=\exp\Bigg [-\bar{K} \sum^{\prime}_{|\sum_{k}(m^k_1- m^k_2)| = {\rm odd}}\Bigg (\prod_k \frac{(\frac{g_k}{\sqrt{N}})^{2(m_1^k+m_2^k)}}{m_1^k!m_2^k!} \Bigg) 
 \int_0^t dt^{\prime}\int_0^{t^{\prime}} dt^{\prime\prime} \big(e^{i\bar{\omega}_mt^{\prime\prime}}+e^{-i\bar{\omega}_m t^{\prime\prime} } \big) \Bigg] \nonumber \\
 &&=\exp\Bigg [-\frac{\bar{K}}{2}  \int_0^t dt^{\prime}\int_0^{t^{\prime}} dt^{\prime\prime}\sum_{\{m_1^k\},\{m_2^k\}}\Bigg\{\prod_k 
\Bigg (\frac{(\frac{g_k}{\sqrt{N}})^{2(m_1^k+m_2^k)}}{m_1^k!m_2^k!}
e^{i{\omega}_k(m_1^k+m_2^k)t^{\prime\prime}} \Bigg)-\prod_k \Bigg (\frac{(-\frac{g^2_k}{N})^{(m_1^k+m_2^k)}}{m_1^k!m_2^k!}
e^{i{\omega}_k(m_1^k+m_2^k)t^{\prime\prime}} \Bigg )\Bigg\}\Bigg ]\nonumber \\
&&~~\times\exp\Bigg [-\frac{\bar{K}}{2} \int_0^t dt^{\prime}\int_0^{t^{\prime}} dt^{\prime\prime} \sum_{\{m_1^k\},\{m_2^k\}}
\Bigg\{\prod_k \Bigg(\frac{(\frac{g_k}{\sqrt{N}})^{2(m_1^k+m_2^k)}}{m_1^k!m_2^k!}
e^{-i{\omega}_k(m_1^k+m_2^k)t^{\prime\prime}} \Bigg)
-\prod_k \Bigg (\frac{(-\frac{g^2_k}{N})^{(m_1^k+m_2^k)}}{m_1^k!m_2^k!}
e^{-i{\omega}_k(m_1^k+m_2^k)t^{\prime\prime}} \Bigg )\Bigg\}\Bigg]\nonumber \\
&&=\exp\Bigg[-\frac{\bar{K}}{2}\int_0^t dt^{\prime}\int_0^{t^{\prime}} dt^{\prime\prime}
\Big[e^{\frac{2}{N}\sum_{k}g_k^2 e^{i\omega_k t^{\prime\prime}}}-e^{-\frac{2}{N}\sum_{k}g_k^2 e^{i\omega_k t^{\prime\prime}}} \Big]\Bigg]
\nonumber \\
&& ~~\times\exp\Bigg[-\frac{\bar{K}}{2}\int_0^t dt^{\prime}\int_0^{t^{\prime}} dt^{\prime\prime}
\Big[e^{\frac{2}{N}\sum_{k}g_k^2 e^{-i\omega_k t^{\prime\prime}}}-e^{-\frac{2}{N}\sum_{k}g_k^2 e^{-i\omega_k t^{\prime\prime}}}\Big]\Bigg]\nonumber \\
&&=\exp\Bigg[-\bar{K}\int_0^t dt^{\prime}\int_0^{t^{\prime}} dt^{\prime\prime}
\bigg[\sinh \bigg(\frac{2}{N\pi}\int_0^\infty \frac{J(\omega)}{\omega^2}e^{i\omega t^{\prime\prime}}d\omega \bigg)
 +\sinh \bigg (\frac{2}{N\pi}\int_0^\infty \frac{J(\omega)}{\omega^2}e^{-i\omega t^{\prime\prime}}d\omega \bigg) \bigg]\Bigg]\nonumber \\
 &&=\exp\Bigg[-2\bar{K}\int_0^t dt^{\prime}\int_0^{t^{\prime}} dt^{\prime\prime}
\cos\bigg(\frac{2}{N\pi}\int_0^\infty \frac{J(\omega)}{\omega^2}\sin\omega t^{\prime\prime}d\omega \bigg)
 \sinh\bigg (\frac{2}{N\pi}\int_0^\infty \frac{J(\omega)}{\omega^2}\cos\omega t^{\prime\prime}d\omega \bigg ) \Bigg].
 \label{multi_diag}
\end{eqnarray}
\end{widetext}
Now, the density of states ${\rm D}(\omega_k)$ for the Einstein model  is given by 
${\rm D}(\omega_k)=N \delta (\omega_k-\omega_0)$ where  N is the  number of phonon modes and $\omega_0$ 
is the fixed frequency for all the modes. Furthermore, in the Einstein model, we can express 
\begin{eqnarray}
 {\rm D}(\omega_k)g_k^2=N \delta (\omega_k-\omega_0)g^2 ,
\label{einst}
\end{eqnarray}
In our case, we are dealing with optical phonons with very weak momentum (k) dependence.
To see the effect of k-dependence for optical phonons, we allow a 
small window for the frequency $\omega_k$ characterized by an upper cutoff frequency $\omega_u$ and a lower cutoff frequency $\omega_l$.
{{Next, we make below a simple generalization of the Einstein-model-based Eq. (\ref{einst}) for our case by replacing the
dirac delta function by a box function of width $\omega_u - \omega_l$ and height $1/(\omega_u - \omega_l)$:}}
{{
\begin{eqnarray}
 {\rm D}(\omega_k)g_k^2&=& g^2 \frac{N}{\omega_u - \omega_l} \Theta (\omega_k - \omega_l) \Theta(\omega_u - \omega_k) ,
 \label{density}
 \end{eqnarray}
where $\Theta(\omega)$ is the unit step function.}}
{{It should be pointed out that for manganites usually $g_k$ is taken as a constant [see  Sec. 3.3.3 of
Ref. \onlinecite{dagotto} and Ref. \onlinecite{held}].}}
 Using the above equation, we calculate the following integrals. 
 \begin{eqnarray}
\frac{1}{N\pi}\int_0^\infty \frac{J(\omega)}{\omega^2} d\omega &=& \frac{1}{N}\sum_k g_k^2 \nonumber \\
&=&\frac{1}{N} \int_0^\infty d\omega_k {\rm D}(\omega_k) g_k^2 \nonumber \\
&=&\int_{\omega_l}^{\omega_u} d\omega_k \frac{g^2}{\omega_u - \omega_l}\nonumber \\
&=&g^2 ,
\label{prefactor}
 \end{eqnarray}
\begin{eqnarray}
\!\!\!\!\!\! \frac{1}{N\pi}\int_0^\infty \frac{J(\omega)}{\omega^2} \cos\omega t^{\prime\prime} d\omega &=& \frac{1}{N}\sum_k g_k^2 \cos\omega_k t^{\prime\prime}\nonumber \\
 &=&\frac{1}{N} \int_0^\infty d\omega_k {\rm D}(\omega_k) g_k^2 \cos\omega_k t^{\prime\prime}\nonumber \\
 &=&\frac{g^2}{(\omega_u - \omega_l)t^{\prime\prime}} (\sin\omega_u t^{\prime\prime} - \sin\omega_l t^{\prime\prime})\nonumber \\
 &=&\frac{2g^2}{(\omega_u - \omega_l)t^{\prime\prime}} \cos\Big[\frac{(\omega_u+\omega_l)t^{\prime\prime}}{2}\Big]\nonumber \\
 &&~~~~~~~~~~~~~~~~\times \sin\Big [\frac{(\omega_u-\omega_l)t^{\prime\prime}}{2}\Big] ,
\nonumber \\
 \label{cos}
\end{eqnarray}
and
\begin{eqnarray}
\!\!\!\!  \frac{1}{N\pi}\int_0^\infty \frac{J(\omega)}{\omega^2} \sin\omega t^{\prime\prime} d\omega &=& \frac{1}{N}\sum_k g_k^2 \sin\omega_k t^{\prime\prime}\nonumber \\
 &=&\frac{1}{N} \int_0^\infty d\omega_k {\rm D}(\omega_k) g_k^2 \sin\omega_k t^{\prime\prime}\nonumber \\
 &=&\frac{g^2}{(\omega_u - \omega_l)t^{\prime\prime}} (\cos\omega_l t^{\prime\prime} - \cos\omega_u t^{\prime\prime}) \nonumber \\
 &=&\frac{2g^2}{(\omega_u - \omega_l)t^{\prime\prime}} \sin \Big [\frac{(\omega_u+\omega_l)t^{\prime\prime}}{2} \Big] \nonumber \\
 &&~~~~~~~~~~~~~~~~\times \sin \Big [\frac{(\omega_u-\omega_l)t^{\prime\prime}}{2}\Big] .
\nonumber \\
 \label{sin}
\end{eqnarray}
Using these expressions, we plot the coherence factor {C(t)} given by equation (\ref{multi_offdiag}) and the inelastic factor P(t) by equation (\ref{multi_diag}) 
in figures (\ref{fig7}) and (\ref{fig8}), respectively.
\begin{figure}
\centerline{\includegraphics[angle=90,angle=90,angle=90,width=3.5in,height=3.5in]{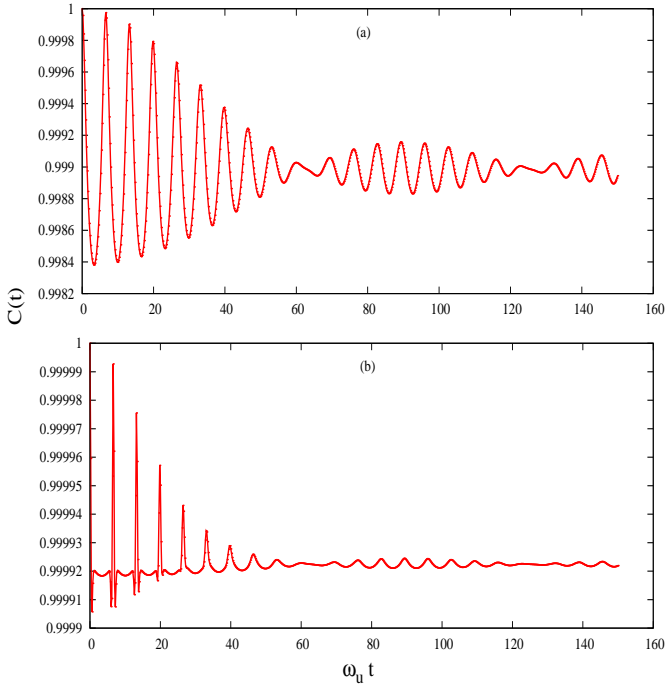}}
\caption{{{Time variation of coherence factor {C(t)} at $\frac{J_\perp}{\omega_u}=.05$ and 
$\frac{\omega_l}{\omega_u}=.9$  for (a) $g=1$ and (b) $g=4$.}}}
\label{fig7}
\end{figure}
Here, we have considered a continuous uniform distribution of $\omega_k$ within a small region characterized by $\omega_l \le \omega_k \le \omega_u$. 
As a consequence of the non-vanishing width $\omega_u - \omega_l$, the harmonics in equations 
(\ref{dfull})--(\ref{dodd}) do not all rephase  at the same time (for $t>0$) since they will always interfere destructively producing no scope
 for complete recoherence (recoherence is only obtained
when one deals with a single phonon frequency).  
 We will now explain the figures (\ref{fig7}) and (\ref{fig8}). Since the upper and the lower cutoff frequencies (i.e., $\omega_u$ and $\omega_l$)
 are quite close to each other, beating effects are seen 
in the figures; the factor $\sin\Big [\frac{(\omega_u-\omega_l)t^{\prime\prime}}{2} \Big]$ [with the beat frequency $(\omega_u-\omega_l)$] 
 produces an envelope for the larger frequency oscillation
 produced by $\cos\Big [\frac{(\omega_u+\omega_l)t^{\prime\prime}}{2}\Big ]$ and $\sin \Big [\frac{(\omega_u+\omega_l)t^{\prime\prime}}{2}\Big ]$
[see equations (\ref{cos}) and (\ref{sin})]. The period of the envelope  
and the faster oscillation are given by $T_{\rm envl}=\frac{2\pi}{\frac{\omega_u-\omega_l}{\omega_u}}=20\pi$ and 
$T_{\rm fast}=\frac{2\pi}{\frac{\omega_u+\omega_l}{2\omega_u}}=\frac{40}{19}\pi$, respectively; thus, one  period  $T_{\rm envl}$
of the envelope contains 
$\frac{19}{2}$ \Big (i.e., $\frac{\frac{\omega_u+\omega_l}{{2}}}{\omega_u-\omega_l}=\frac{19}{2}$ \Big ) faster oscillations. 
At small values of $\omega_u t^{\prime\prime}$, the nature of the plot is similar to the case 
when only one single phonon mode $\omega$ is considered. The beating effect is not observed at smaller time as 
$\lim_{\omega_u t^{\prime\prime}\rightarrow 0}\frac{\sin \frac{(\omega_u-\omega_l)t^{\prime\prime}}{2}}{\frac{(\omega_u-\omega_l)t^{\prime\prime}}{2}}=1$
 and we get oscillations with frequency $\frac{\omega_u+\omega_l}{2}\approx\omega_u$. 
As $\omega_u t^{\prime\prime}$ increases from zero value, the factor 
$\frac{\sin \frac{(\omega_u-\omega_l)t^{\prime\prime}}{2}}{\frac{(\omega_u-\omega_l)t^{\prime\prime}}{2}}$ decreases 
and contributes very little at values of $\omega_u t^{\prime\prime}$ beyond $40\pi$. Consequently, the amplitude of the oscillations, 
in the exponent of the equations (\ref{multi_offdiag}) and 
(\ref{multi_diag}), decreases rapidly as $\omega_u t^{\prime\prime}$ increases from zero value.
{{In the oscillations of figures (\ref{fig7}) and (\ref{fig8}), this describes the damping
  due to the
  non-vanishing width of the frequency distribution.}}
\begin{figure}
\centerline{\includegraphics[angle=90,angle=90,angle=90,width=3.5in,height=3.5in]{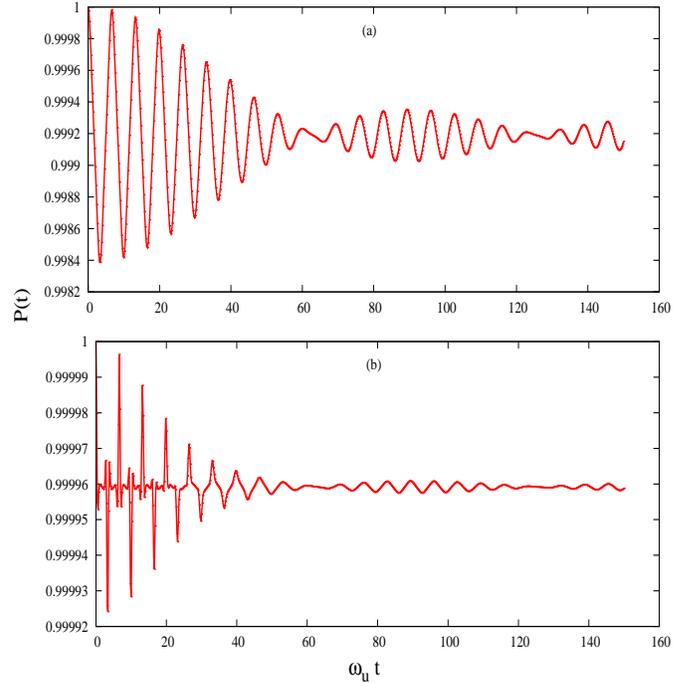}}
\caption{Time variation of inelastic factor P(t) at $\frac{J_\perp}{\omega_u}=.05$ and $\frac{\omega_l}{\omega_u}=.9$ for (a) $g=1$ and (b) $g=4$.}
\label{fig8}
\end{figure}
}}

\end{document}